\documentclass{article}
\pdfoutput=1
\RequirePackage[OT1]{fontenc}
\RequirePackage{amsmath,amsfonts,amssymb,amsthm}
\RequirePackage{natbib}
\usepackage{authblk}
\RequirePackage[colorlinks,citecolor=blue,urlcolor=blue]{hyperref}

\usepackage[toc,page]{appendix}

\usepackage[normalem]{ulem}
\usepackage{graphicx,bookmark}
\usepackage{booktabs,mathtools}
\usepackage{rotating}
\usepackage{here}
\usepackage[margin=1.25in]{geometry}
\usepackage{dsfont}
\usepackage{booktabs}
\usepackage{multirow}
\numberwithin{equation}{section}

\newtheorem{theorem}{Theorem}[section]

\theoremstyle{remark}
\newtheorem{definition}[theorem]{Definition}
\newtheorem*{example}{Example}

\begin{document}
\title{On foundation of generative statistics with G-entropy: a gradient-based approach}

\author[1]{Bing Cheng}
\author[2]{Howell Tong}

\affil[1]{\small Academy of Mathematics and System Science, Chinese Academy of Sciences,  Haidian District, Beijing 100190, China}
\affil[2]{Department of Statistics and Data Science, Tsinghua University, Beijing 100084, China; Department of Statistics, London School of Economics and Political Science , London WC2A 2AE, UK.}

\date{\small \today}

\maketitle


\begin{abstract}
This paper explores the interplay between statistics and generative artificial intelligence. Generative statistics, an integral part of the latter, aims to construct models that can {\it generate} efficiently and meaningfully new data across the whole of the (usually high dimensional) sample space, e.g. a new photo. Within it, the gradient-based approach is a current favourite that exploits effectively, for the above purpose, 
the information contained in the observed sample, e.g. an old photo. However, often there are missing data in the  observed sample, e.g. missing bits in the old photo. To handle this situation, we have proposed a gradient-based algorithm for generative modelling. More importantly, our paper underpins rigorously this powerful approach by introducing a new G-entropy that is related to Fisher's divergence. (The G-entropy is also of independent interest.) The underpinning has enabled the gradient-based approach to expand its scope. For example, it can now provide a tool for generative model selection. Possible future projects include discrete data and Bayesian variational inference.
\end{abstract}

\section{Introduction}

Generative artificial intelligence (AI) has been hailed as signaling the  beginning of the 4th industrial revolution.
As an integral part of  it, generative statistics has the following characteristics:
(1) It deals with a data distribution $p_{data}(x)$ (equivalently denoted as $p_x$) and  a model distribution $p_{model}(x,\theta )$ that are 
complex and live in a high dimensional data space $S\subset R^d$. In generative AI, the dimension of $S$ could range from hundreds to thousands and that of the parameter space, $\Theta$, from millions to trillions. (2) It finds ways to approximate  a high dimensional $p_x$  by an appropriate $p_{model}(x,\theta )$. 
(3) It finds  efficient and flexible (the so-called human prompting) ways to generate new samples $x_1,\cdots, x_n$ from $p_{model}(x,\theta )$. 
\footnote {For convenience, we shall use the lower case to denote both the random variable and its realized value. The text should make it clear as to which is intended. Furthermore, we shall focus on probability density functions (PDF) in this paper. We comment on discrete distributions in Subsection 4.3.}
(4) It handles  the relationship between its capacity and the number of parameters, the complexities, the size of the training data and the computation power (the well-known  Scaling Law). (5) Its versatility in handling
  data from vastly different areas, logical reasoning, common-sense reasoning, art of creativity and others, is achieved not so much by a conventional specific series of models but rather by  a single generative model via various professional prompts. (6)  No generative models can be free from  hallucination:  incorrect or misleading results can be generated.  

  A primary focus of conventional statistics is to {\it analyse} the observed data that typically live in a subset of $S.$ Contrarily, a focus  of generative statistics is to {\it generate} new, more relevant and meaningful data samples across typically high-dimensional $S,$ often  involving high dimensional $\Theta$, hence the adjective {\it generative}: be creative in a meaningful way. This is the very challenge taken on by the generative statistics.

  Now, the century-old maximum likelihood method is directly linked to the Jeffreys-Kullback-Leibler (JKL)
\footnote{Jeffreys predated Kullback and Leibler when he introduced the above divergence on page 158 in his book {\it Theory of Probability} (second edition), Oxford University Press, 1948.} divergence that seeks {\it point-wise} closeness between  $p_{data}(x)$ (or simply $p_x$) and $p_{model}(x,\theta )$. However, it is highly infeasible to check the point-wise closeness within the whole of $S$. Current ways to overcome this difficulty involve focusing on directions of the log-likelihood functions, namely gradient vectors or gradient vector fields of $p_x$ and $p_{model}(x,\theta )$ {\it with respect to $x.$}  Among these gradient-based methods are the Stein's identity (1981), Hyvärinen's score-matching method (2005) and others. An advantage of these methods is the facility in bypassing intractable partition functions. Furthermore, the Langevin dynamics, or a general diffusion process, exploits the information of the gradient vectors to generate new data efficiently in $S$. In fact, gradient-based methods are particularly useful in handling {\it implicit} 
distributions, namely those that are intractable (e.g. posterior distributions with no closed forms) but from which tractable processes (e.g. MCMCs) exist to sample. However, there are often missing data in the observed sample. We have developed a new gradient-based algorithm to handle them.

There exists a natural connection between the JKL divergence and the Shannon entropy that studies statistical information of a log-likelihood function and to which the maximum likelihood methodology is intimately linked. It is, therefore, relevant to ask  { \it'Which information notion underpins the gradient-based methods?'} To address this fundamental question, we start, in this paper, from a different divergence, the Fisher divergence, which measures the closeness between gradient vectors or gradient vector fields. Based on it, we introduce a new  entropy, called the G-entropy, to measure gradient information of a log-likelihood function. The G-entropy is of independent interest. Quite remarkably,  
it possesses all the properties of the Shannon entropy, except for the convexity property. Importantly, once underpinned by the G-entropy, the gradient-based approach can be expanded  to address  issues such as generative model selection.

\subsection{Noise and diffusion models}

A major advance from the original score matching method (to be described later) in generative modelling is by shifting to a corrupted sample by incorporating a noise term $\epsilon_t.$ However, serious difficulties remain, because what is generated is the corrupted data $\tilde {x}$ from  $p_{\sigma}(\tilde{x})$ and not $p_x$. To mitigate this situation,  a series of corrupted samples from $p_{\sigma_i}$ with 
	\[
	\sigma_{max}=\sigma_1 > \sigma_2 > \cdots > \sigma_{min}=\sigma_T > 0,
	\]
	(-the so-called noise schedule)
	was proposed in the hope that $p_{\sigma_{min}}\approx p_x$. 

 Most currently popular and efficient generative models possess two components:
	\begin{itemize}
		\item \textbf{modelling component-} streamlined information embedded in the sample data $x$ from $p_x$, often associated with a deep encoder such as a transformer in chatGPT and variational autoencoders (VAEs);
		\item \textbf{sampling component-} a model or a procedure, often  associated with a deep decoder, Langevin dynamics or MCMC, to generate a new sample $x_{new}$ whose distribution is close to $p_x$,

	\end{itemize}
	
	There is a parent-son relationship in that the generating functions used in the sampling component are to come from the modelling component, a task of training a generative model.
	In this regard, we can see the potential  of the noise schedule: 
 \begin{equation}
		x_{t+1}=x_t + {\epsilon_{t+1}}, 
  \epsilon_{t+1} \sim N(0,\sigma_{t+1}) \text{ for } t=0,1,\cdots, T,
	\end{equation}
	with a suitably chosen $x_0=x$.
 When $T$ is sufficiently large, the corrupted sample is almost a white noise with no  structure left.
	From the above equation 
, we can generate the original sample  from the white noise $x_T$ by a reverse process or de-noise process. For example, the Langevin dynamics gives
	\[
	x_t=x_{t+1} + \frac{\delta}{2}\nabla_x log~p(x_{t+1}) + \sqrt{\delta}{z_t} = x_{t+1}+e_{t+1},
	\]
	where $\delta$ is the step size and $z_t$ is a standard Gaussian random variable, and the score function of the noisy sample, $\nabla_x log~p(x_{t+1})$, is estimated at the modelling stage, also known as the forward process. Now, the noise distribution tends to be more evenly spread with minimal danger f suffering from the manifold hypothesis; another new idea is introduced to the diffusion model, namely the focus on the score function of the noise rather than that of the data. For illustrations of the gradient-based image generating process - the so-called diffusion models in AI,
 we refer to https://lilianweng.github.io/posts/2021-07-11-diffusion-models/.

\section{Gradient-based generative modelling}

\subsection{Gradient-based models: Score-matching method}

Let $x$ be a random vector in $R^d,~d\geq 1$ with PDF $p_{data}(x)$.
 Let $x_1,\cdots ,x_n$ be a random sample from $p_{data}(x)$. The  score function
 \begin{equation}
		s(x)=\nabla_x \log p_{data}(x) 
	\end{equation} 
 steers a point $x$ in the direction of increasing 
	$\log p_{data}(x)$  to the next  point $x+\lambda s(x).$ 
	
	However, given an unknown $p_{data}(x)$, Hyv{\"a}rinen (2005) suggested approximating it with  a parametric PDF, namely $p_{model}(x,\theta)$, exemplified in the energy functional form.  To measure the discrepancy between their scores, he used the the Fisher divergence $J(\theta)$:
 \[
	J(\theta)=\frac{1}{2}E_{p_{data}}||s(x,\theta)~-~s(x)||^2,
	\]
	\begin{equation}
		~=~\frac{1}{2}E_{p_{data}}||\nabla_x \log p_{model}(x,\theta)~-~\nabla_x \log p_{data}(x)||^2,
	\end{equation}
	by defining $s(x,\theta)=\nabla_x \log p_{model}(x,\theta)$. 
	He called this method the {\it score matching} method, with which the unknown parameter $\theta$ can be obtained by minimizing the Fisher divergence, i.e. \begin{equation}
		\theta^*~=~arg \min_{\theta \in \Theta}\{ J(\theta)\}.
	\end{equation}
	To generate new samples in the data space, we move in the direction guided by $s(x,\theta^*)$  starting from any random point $x_{0}$ over a finite time horizon. In practice, we move with a little bit of stochasticity, under the name of {\it Langevin Dynamics}:
	\begin{equation}
		x_{t+1}=x_t+\alpha_t\times s(x_t,\theta^*)~+~\sqrt{2\alpha_t}z,
	\end{equation}
	where z is the injected Gaussian noise, namely $z\sim N(0,I)$. If $\alpha_t \rightarrow 0$ as $t\longrightarrow \infty$, this process guarantees that the limit of $x_t$ is a true sample from $p_{model}(x,\theta^*)$. In practice, we run this process for a finite number of steps 
	and assign $\alpha_t$ according to a decaying schedule. See Welling {\it et al.} (2011). 
	Note that we will not get $\theta^*$ since 
	$p_{data}(x)$ 
		is unknown. 

\subsection{Fisher divergence and the objective function $W$}

Let $p(x)$ stand for a generic PDF
 on $R^d$ that satisfies the following assumptions: 
	\begin{itemize}
		\item {\bf assumption 1:} $p(x)$ is  twice differentiable 
		on $R^d$;
  \item {\bf assumption 2:} $p(x)$, 
		$\nabla_x p(x)$ and 
		$\nabla_x \nabla_x p(x)$ are all square-integrable on $R^d$;
		\item {\bf assumption 3:} for every $x\in R^d$ with $x=(x_1,\cdots ,x_d)$ and  for each boundary point of $x_i,~i=1,\cdots ,d$, 
		\[
		p(x_1,\cdots,x_{i-1},-\infty ,x_{i+1},\cdots ,x_d)\equiv 0 \text{ and } p(x_1,\cdots,x_{i-1},+\infty ,x_{i+1},\cdots ,x_d)\equiv 0.
		\]
	\end{itemize}

 Henceforth, all PDFs are assumed to satisfy assumptions 1 and 2.
 \begin{definition}
		The Fisher divergence between PDF $p$ and PDF $q$ is defined as
		\begin{equation}
			D_F(p||q)=\frac{1}{2}E_p||\nabla_x \log p(x)~-~\nabla_x \log q(x)||^2.
		\end{equation}
		Note that it  
		is non-negative, not symmetric,
		zero if and only if p and q are equal p-almost everywhere, and  does not depend on the normalizing constants associated with p and q. Note also that $J(\theta) = D_F(p_{data}(x) || p_{model}(x,\theta))$. 
\end{definition}
	
	\begin{definition}
 {\it The objective function} $W$ of PDF $q$ at point $x$ is defined by
		\begin{equation}
			W(x,q)=-||\nabla_x \log q(x)||^2~-~2\Delta_x \log q(x),
		\end{equation}
  where 
		$\Delta$ denotes the Laplacian, {\it i.e. } $\Delta_x f(x) = \sum_{i=1}^d \frac{\partial^2 f(x)}{\partial^2 x_i}$.
	\end{definition}
	
Note that $-W$ is just the Hyvarinen score (Hyvarinen (2015).) 
	
{\bf Theorem 1}:
		Let p and q be two PDFs on $R^d$ and let p satisfy additionally assumption 3. We have
		\begin{equation}
			E_p||\nabla_x \log p(x)~-~\nabla_x \log q(x)||^2~=~E_p||\nabla_x \log p(x)||^2~-E_p W(x,q).
		\end{equation}
	We give a proof in the Supplementary Material.
 	
	From the above theorem 
we have immediately 

{\bf Corollary 1}:

		\begin{equation}
			\theta^*=arg \min_{\theta \in \Theta} J(\theta)~=~arg \max_{\theta \in \Theta}E_{p_{data}}\{W(x,p_{model}(x,\theta))\}.
		\end{equation}

\subsection{Fisher divergence and Jeffreys-Kullback–Leibler divergence}

	
	Recall that the JKL divergence   between PDFs $p$ and $q$ is defined by 
	\begin{equation}
		D_{JKL} (p||q) = E_p \log \{p(x)/q(x)\}.  
	\end{equation}
	
	Lyu (2009)
 proved the following interesting result between the Fisher divergence and the JKL divergence
	\begin{equation}
		\frac{d}{dt}D_{JKL} 
		(\tilde{p}_t||\tilde{q}_t)~=~-\frac{1}{2}D_F 
		(\tilde{p}_t||\tilde{q}_t)),
	\end{equation}
	where typically $\tilde{p}_t$ denotes the pdf of $\tilde{x}$ that is a corrupted version of $x$  given by
	\[
	\tilde{x} = x + \sqrt{t}\epsilon, ~\text{with } x \sim p ~\text{and  }  \epsilon \sim N(0,I).
	\]
	
	
	And, in particular,
	\begin{equation}
		\frac{d}{dt}D_{JKL} (\tilde {p}_t||\tilde{q}_t)|_{t=0}~=~-\frac{1}{2}D_F(p||q).
	\end{equation}
 
\subsubsection{Challenges faced by the Score Matching Method}

Song and Ermon (2019)  
 addressed fundamental hurdles for a naive score 
	matching method stemming from the so-called {\it manifold hypothesis}, which states that  data in the real world tend to be concentrated on low dimensional manifolds embedded in a high dimensional data space. Apparently, it holds empirically for many datasets, and is of  foundational interest in manifold learning. 
	Under it,
	score-based generative models face two key difficulties. First,  the score function
	is undefined when the data are confined to a low dimensional manifold. Second, a consistent estimator of the objective function is possible 
	only when the support of the data distribution is the whole space (Theorem 2 in Hyvärinen  (2005)).

\section{Foundation of gradient-based modelling: entropy, divergence and information}

\subsection{Shannon entropy}

The Shannon approach
is based on the following 
objective function (Shannon objective function/functional)
\begin{equation}
OB(x,p)=-\log p(x),
\end{equation}
from which
a series of 
concepts have been introduced, e.g. entropy, 
cross-entropy, joint entropy, conditional entropy,   and others, which are available in many standard text-books. Of particular relevance here is the  following obvious link between the JKL divergence and the Shannon entropies. Consider PDFs $p$ and $q$. By definition, 

\begin{equation}D_{JKL} (p||q)= E_{p}[OB(x,q)]-E_{p}[OB(x,p)] = H(p,q) - H(p),
\end{equation}
where $H(p,q)=E_{p}[OB(x,q)]$ is known as the Shannon cross-entropy, and $H(p) = E_{p}[OB(x,p)]$ is known as the Shannon entropy. 

\subsection{G-entropy and information}

Although the  maximum likelihood principle, the Shannon entropy and JKL divergence, all inter-related, are popular and intuitive approaches, their success rests mostly with cases of tractable PDFs. Essentially, they focus on the pointwise information provided  by the PDF.  For intractable PDFs, in particular implicit PDFs, they face serious challenges. 
As an alternative, we  adopt  in this paper an approach
based on the {\it gradients} of the PDFs and the objective function $W$. It parallels 
the above approaches and is better prepared to address issues with implicit PDFs. First, we introduce entropies that parallel Shannon's.

(1) G-entropy:

\begin{equation}
	H_G(x)=H_G(p)=E_p||\nabla_x\log p(x)||^2.
\end{equation}

(2) G-cross-entropy between PDF $p$ of random variable $x$ and PDF $q$ of $y$:
	\begin{equation}
		H_G(p,q)=E_p[W(x,q)].
	\end{equation}

(3) G-joint-entropy:
 	
	For random vectors $x$ and $y$ of possibly different dimensions, let $p_{(x,y)}(x,y)$ be the joint PDF  of $(x,y)$. The G-joint-entropy of x and y is defined as
	\begin{equation}
		H_G(x,y)=H_G(p_{(x,y)}).
	\end{equation}

(4) G-mutual information: 

Let $x$ and $y$ be two random vectors. The G-mutual information is defined as 
\footnote{The Shannon mutual information is defined as $I(x,y)=H(x)+H(y)-H(x,y).$}
	\begin{equation}
		I_G(x,y)=H_G(x,y)-H_G(x)-H_G(y).
	\end{equation}

(5) G-conditional entropy:  

Let $x$ and $y$ be two random vectors with joint PDF  $p(x,y)$. Let $p(x)$ be the PDF of $x$. The F-conditional entropy of $y$ given $x$ is defined as
	\begin{equation}
		H_G(y|x)=E_{p(x,y)}[W((x,y),\frac{p(x,y)}{p(x)})].
	\end{equation}
	
(6) G-partial information: 

For $i=1,\cdots, d$, we define the G-partial information in the direction of the $x_i$ axis  in the data space by
	\begin{equation}
		GI_i=E_p[\frac{\partial \log p(x)}{\partial x_i}]^2.
	\end{equation} 
	
(7) G-information matrix:

	Let $x$ be a random vector. We define the G-information matrix $GIM_x$ by
\footnote{We might simply write GIM when the context is clear.} 
	\[
	GIM_x=E_p[[\nabla_x\log p(x)] [\nabla_x\log p(x)]^T].
	\]

By the equation in Definition 2 and Theorem 1, 
we have 

\begin{equation}
D_F(p_{(x,y)}||p_xp_y)=\frac{1}{2}\{H_{G}(p_{(x,y)})-E_{p_{(x,y)}}W(\text{\small (x,y)},~p_xp_y)\}.
\end{equation}

{\bf Proposition 1}:
Under assumptions 1-3 for p, we have
\begin{enumerate}
	\item 
	
	\begin{equation}
		E_p[\nabla_x\log p(x)]~=0;
	\end{equation}
	\item For $x=(x_1, \cdots,x_d)$,
	\begin{equation}
		GI_i=E_p[\frac{\partial \log p(x)}{\partial x_i}]^2~=~-E_p[\frac{\partial^2 \log p(x)}{\partial^2 x_i}]=Var[\frac{\partial \log p(x)}{\partial x_i}],~i=1,\cdots ,d.
	\end{equation}
	\item 
	\begin{equation}
		H_G(p)=-E_p[\Delta_x \log p(x)].
	\end{equation}
\end{enumerate}
We give a proof in  the Supplementary Material.

From Proposition 1, the following proposition is immediate.


{\bf Proposition 2}:
 
	\begin{enumerate}
		\item The G-entropy is an uncertainty measure of the 
		score functions:
		\begin{equation}
			H_G(p)=\sum_{i=1}^d Var(\frac{\partial \log p(x)}{\partial x_i})=-E_p[\Delta_x \log p(x)].
		\end{equation}
		\item The G-entropy is the sum of the G-partial information or the trace of the G-information matrix, i.e.
		\begin{equation}
			H_G(p)=\sum_{i=1}^dGI_i=trace(GIM).
		\end{equation}
	\end{enumerate}
	

{\bf Proposition 3}:
When $q=p$, we have 
\begin{equation}
	H_G(p,q)=H_G(p)=E_p[W(x,p)].
\end{equation}

\begin{proof}
By the definition of the G-cross-entropy, we have
\[
H_G(p,p)=E_p[W(x,p)]=-E_p||\nabla_x \log p(x)||^2-2E_p[\Delta_x \log p(x)].
\]
However,
\[
E_p||\nabla_x \log p(x)||^2=\sum_{i=1}^dE_p[\frac{\partial \log p(x)}{\partial x_i}]^2
\]
\[
=-\sum_{i=1}^dE_p[\frac{\partial^2 \log p(x)}{\partial^2 x_i}]=-E_p[\Delta_x \log p(x)]
\]
by Proposition 1.
Hence, by the definition of  the G-entropy and equation (3.12) 
it holds that
\[
H_G(p,p)=E_p[\Delta_x \log p(x)]-2E_p[\Delta_x \log p(x)]=-E_p[\Delta_x \log p(x)]=H_G(p).
\]

The proof is complete. 
\end{proof}
A similar relationship holds with the Shannon entropy, namely $H(p,p) = H(p)$.

{\bf Corollary 2}:
For any q and p, we have
\begin{equation}
	H_G(p) \geq H_G(p,q).
\end{equation}
That is, given p, 
\begin{equation}
	\max_{q}\{H_G(p,q)\}=H_G(p).
\end{equation}

\begin{proof}
	By Proposition 3,
we have $H_G(p,p)=H_G(p)$ and then by Theorem 1
 and the non-negativity of Fisher divergence, we obtain the inequality for any given q.
\end{proof}
By interpreting p as a data distribution and q as a model distribution, we obtain the best approximation of p by q by maximizing the G-cross-entropy. We call this {\bf the G-entropy maximization principle}, which is reminiscent of {\it Akaike's entropy maximization principle} (e.g., Akaike (1985)) that led to the well-known AIC. Later, we shall show how the the G-entropy maximization principle leads to an GIC. 

{\bf Proposition 4}:
Let x be a random vector in $R^d$ with  PDF $p_x(x)$ and  y be a random vector in $R^e$ with PDF  $p_y(y)$. Let $p_{(x,y)}(x,y)$ be the joint PDF  of $(x,y)$. Then under assumptions 1 and 2 for $p_x(x)$ and $p_{(x,y)}(x,y)$, we have
\begin{equation}
	E_{p_{(x,y)}}||\nabla_x \log p_x(x)]||^2=H_G(p_x).
\end{equation}
\begin{proof}
\[
E_{p_{(x,y)}}||\nabla_x \log p_x(x)||^2=\sum_{i=1}^d\int_x\int_y [\frac{\partial \log p_x(x)}{\partial x_i}]^2 p_{(x,x)}(x,x)dxdy
\]
\[
=\sum_{i=1}^d\int_x \{[\frac{\partial \log p_x(x)}{\partial x_i}]^2 \int_y p_{(x,y)}(x,y)dy\}dx
\]
\[
=\sum_{i=1}^d\int_x \{[\frac{\partial \log p_x(x)}{\partial x_i}]^2 p_x(x)\}dx=\int_x||\nabla_x\log p_x(x)||^2 p_x(x)dx=H_G(p_x).
\]
\end{proof}

{\bf Theorem 2}:
Under assumptions 1 and 2, the following relationship between the Fisher divergence and the G-mutual information holds for x and y as defined in Proposition 4:
\begin{equation}
	D_F(p_{(x,y)}||p_xp_y)=\frac{1}{2}I_G(x,y).
\end{equation}
We give a proof in the Supplementary Material. 

The following corollaries are immediate. 

{\bf Corollary 3}:
The G-mutual information is non-negative
\begin{equation}
	I_G(x,y)\geq 0
\end{equation}

{\bf Corollary 4}:
If x and y are independent, then the G-mutual information is zero
\begin{equation}
	I_G(x,y)=0
\end{equation}
and the G-joint-entropy is equal to the sum of two individual G-entropies, i.e.
\begin{equation}
	H_G(x,y)=H_G(x)+H_G(y).
\end{equation}

{\bf Lemma 1}:
For x and y as defined in Proposition 4,
\begin{equation}
E_{p_{(x,y)}}[[\nabla_x\log p(x,y)]^T[\nabla_x \log p(x)]]=-E_{p_x}[\Delta_x\log p(x)].
\end{equation}
We give a proof  in the Supplementary Material.

{\bf Theorem 3}:
For x and y as defined in Proposition 4, the G-conditional entropy satisfies 
\begin{equation}
	H_G(y|x)=H_G(x,y)-H_G(x).
\end{equation}
We give a proof  in the Supplementary Material.

\vspace{0.5cm}
{\bf Corollary 5}:

If x and y are independent, then
\begin{equation}
	H_G(y|x)=H_G(y)
\end{equation}
and
\begin{equation}
	H_G(x,y)=H_G(x)+H_G(y).
\end{equation}

\begin{proof}
By definition, $H_G(y|x)=E_{p(x,y)}[W((x,y),\frac{p(x,y)}{p(x)})].$ By independence, $p(x,y)=p(x)p(y)$. Therefore,
\begin{equation}
	E_{p(x,y)}[W((x,y),\frac{p(x,y)}{p(x)})] = E_{p(y)}[W(y, p(y))] = H_G(p(y)).
\end{equation}

\end{proof}
\begin{example}
To illustrate the various notions, 
let x be a $d\times 1$ multivariate 
random vector with PDF $N(\mu, \Sigma)$, mean vector  $\mu$ and variance-covariance matrix $\Sigma$. We have

\begin{equation}
	\nabla_x \log p(x)=\nabla_x [-\frac{1}{2}(x-\mu)^{\tau}\Sigma^{-1}(x-\mu)]=-\Sigma^{-1}(x-\mu),
\end{equation}
where 
$\Sigma^{-1}$ is the inverse of $\Sigma$. We can easily check that

\begin{equation}
	H_G(x)=trace (\Sigma^{-2} \Sigma)=trace(\Sigma^{-1}).
\end{equation}

Further, suppose that $x_1,\cdots ,x_d$ are independent  with  $x_i\sim N(\mu_i,\sigma_i)$. Then 
\[
H_G(x)=H_G(x_1,\cdots ,x_d)=\frac{1}{\sigma^2_1}+\cdots+\frac{1}{\sigma^2_d}=\frac{d}{\sigma^2} 
\] 
if the variances are all equal to $\sigma^2$.

For the special case of $d=2$ with $x_1 = x$, $x_2 = y$ and 
\[
\Sigma=	\begin{bmatrix}
	\sigma_{x}^2 & \sigma_{x,y} \\
	\sigma_{y,x} & \sigma_y^2
\end{bmatrix},
\]
we have
$$H_G(x,y)=\frac{1}{(1-\rho^2) }[\frac{1}{\sigma_x^2} + \frac{1}{\sigma_y^2}],$$
\begin{equation}
	I_G(x,y) = \frac{\rho^2}{1-\rho^2}[\frac{1}{\sigma_x^2} + \frac{1}{\sigma_y^2}].
\end{equation}
If x and y are independent, then $H_G(x,y) = H_G(x) + H_G(y)$, and $I_G(x,y)=0$. 

Next,

\[
H_G(y|x)=H_G(x,y)-H_G(x)=\frac{1}{1-\rho^2}[\frac{1}{\sigma_x^2} + \frac{1}{\sigma_y^2}]-\frac{1}{\sigma_x^2},
\]

and \[
H_G(p_x,p_y)=
[2-\frac{\sigma_x^2}{\sigma_y^2}-\frac{(\mu_x-\mu_y)^2}{\sigma_y^2} ]\times FI_y,
\]
where $GI_y = \frac{1}{\sigma_y^2}$ is the G-partial information of $y$.   
Finally,  for any given $p_x$,
maximizing $H_G(p_x,p_y)$ with respect to $p_y$ gives  $\mu_y = \mu_x$ and  $\sigma_y^2=  \sigma_x^2$ as expected from Corollary 2. 
\end{example}

Sueli  {\it et al.} (2015)
studied the geometric analysis of the F-information matrix with respect to the parameters of a univariate Gaussian PDF as well as  to $x$ in a data space $R^d$.

\subsection{Applications of Fisher divergence, entropy and information}

\subsubsection{Bayes variational approximation by Fisher divergence-Fisher variational approximation}

Let $p(y|x)$ be a posterior distribution. Then we use a tractable distribution $q(y)$ to approximate $p(y|x)$.  The $q(y)$ is called  the {\it variational distribution}. It aims to be similar to the true posterior and can be chosen from a family of simpler distributions, like the Gaussian distribution or other tractable ones (e.g.,
Kingma and Welling (2022)). JKL divergence and the evidence lower bound (ELBO) are two often used measures of the discrepancy between $p(y|x)$ and $q(y)$.

As another method, let $p(y)$ be a distribution (tractable or intractable). MCMC (in particular, the Metropolis-Hastings algorithm) is used to construct a Markov chain with conditional transition distribution $q(y_t|y_{t-1})$, such that the sample $y_t$ generated by this Markov chain will converge to that governed by the distribution $p(y)$. 

The two methods are popular but each has its own limitations:\begin{itemize}\item {\bf Speed and Scalability:} MCMC can be slow, especially for complex models and high accuracy, especially in high dimensional spaces. Bayes variation is much faster, especially for high-dimensional problems. However, the accuracy of Bayes variational approximation depends on the chosen family and the optimization.
\item {\bf Approximation of moments:} 
Song {\it et al.} (2019)
pointed out that, by minimizing  KL divergence between $p(y|x)$ and $q(y)$, the Bayes variational method provides an accurate and efficient estimation of the {\bf posterior mean}, but often failing to capture other moments, and has limitations regarding to which models it can be applied. 
\item {\bf Direct sampling}: The Bayes variational method does not provide sampling but estimator of the posterior.  
\end{itemize}

On comparison with other methods like the mean-field variational inference, the results of Yang {\it et al.} (2019) demonstrate that a Fisher divergence-based variational approximation is substantially better in terms of the posterior mean and  covariance matrix.

\subsubsection{Consistency and asymptotic normality}

Let $p_{M}(\theta)$ be a model PDF with unknown $h$-dimensional parameter vector $\theta$ ($h\geq 1$). Let $x_1,\cdots ,x_n$ be a random sample in $R^d$ from the data PDF $p_x$. We estimate  $\theta$ as follows.

\begin{equation}
    \hat{\theta}=\hat{\theta}_n=arg \max_{\theta \in \Theta}\{ GIC(M(\theta))\},
    \end{equation}
    where $GIC(M(\theta))=\frac{1}{n}\sum_{i=1}^n 
	W(x_i,p_{M(\theta)})$. We call $\hat{\theta}$ the maximum GIC estimate (MGICE) of $\theta$. Note that for the multivariate Gaussian PDF $N(\bf\mu,\Sigma)$, the MGICEs are the same as their maximum likelihood estimate (MLE) counterparts, although this is not generally the case. For example, MGICE is not defined for the exponential PDF  
 because 
 it violates Assumption 5 below.

Song {\it et al.} (2019) have proved the consistency and asymptotic normality for sliced score-matching estimator. Here we will use their results under a special projection matrix $v$. First we list the assumptions that they have used as follows.
\begin{itemize}
\item {\bf Assumption 4:} 
$p_x=p_{M(\theta^*)},$ where $\theta^*$ is the true parameter in $\Theta$. Furthermore, $p_{M(\theta)} \neq p_{M(\theta^*)} $ whenever $\theta \neq \theta^*$; this is one of the standard assumptions used for proving the consistency of MLE as well.
\item {\bf Assumption 5:} 
$p_{M(\theta)}(x) > 0,~\forall \theta \in \Theta$ and $\forall x$. This is the assumption used in Hyvärinen (2005).
\item {\bf Assumption 6:} The parameter space $\Theta$ is compact. This is a standard assumption used for proving the consistency of MLE.
\item {\bf Assumption 7:}  Both $\nabla_x^2\log p_{M(\theta)}(x)$ and $[\nabla_x\log p_{M(\theta)}(x)][\nabla_x\log p_{M(\theta)}(x)]^T$ are Lipschitz continuous in respect of
Frobenious norm. Specifically, $\forall \theta_1,\theta_2 \in \Theta$, 
\[
||\nabla_x^2\log p_{M(\theta_1)}(x)-\nabla_x^2\log p_{M(\theta_2)}(x)||_F\leq L_1(x) ||\theta_1-\theta_2||_2, 
\]
and
\[
||[\nabla_x\log p_{M(\theta_1)}(x)][\nabla_x\log p_{M(\theta_1)}(x)]^T-[\nabla_x\log p_{M(\theta_2)}(x)][\nabla_x\log p_{M(\theta_2)}(x)]^T||
\]
\[\leq L_2(x) ||\theta_1-\theta_2||_2. 
\]

In addition, $E_{p_x}[L_1^2(x)]<\infty$ and $E_{p_x}[L_2^2(x)]<\infty.$
\item {\bf Assumption 8:}  (Lipschitz smoothness on second derivatives).

For $\theta_1$, $\theta_2$ near $\theta^*$, and $\forall i, j$,
\[
||\nabla_{\theta}^2 \partial_i \partial_j \log p_{M(\theta_1)} - \nabla_{\theta}^2 \partial_i \partial_j \log p_{M(\theta_2)}||_F \leq M_{i,j}(x)||\theta_1 - \theta_2||_2
\]
and
\[
||\nabla_{\theta}^2 \partial_i \log p_{M(\theta_1)} \partial_j \log p_{M(\theta_1)}- \nabla_{\theta}^2 \partial_i \log p_{M(\theta_1)} \partial_j \log p_{M(\theta_2)}||_F \leq N_{i,j}(x)||\theta_1 - \theta_2||_2.
\]
Here, $\partial_i$ refers to  the partial derivative 
with respect to the component $x_i$ in the random vector $x=(x_1,\cdots ,x_d).$

\end{itemize}

\vspace{0.5 cm}
{\bf Proposition 5}: 
{Let $\theta^*$ be the true parameter of $p_x$. Under Assumptions 1-7, we have 
    \begin{equation}
        \hat{\theta}_n \stackrel{p}{\longrightarrow} \theta^*~\text{ as }n\longrightarrow \infty.
    \end{equation}

We give a proof in the Supplementary Material.

{\bf Lemma 2}:
Suppose that $p_{M(\theta)}(x) $ is sufficiently smooth (Assumption 8). Then $\nabla^2_{\theta}FIC(M(\theta))$ is Lipschitz continuous, i.e., for $\theta_1$ and $\theta_2$ close to $\theta^*$, there exists a Lipschitz constant $L(x)$ such that
    \begin{equation}
        || \nabla^2_{\theta}GIC(M(\theta_1)) - \nabla^2_{\theta}GIC({M(\theta_2)})||_F \leq L(x)||\theta_1 - \theta_2||_2.
    \end{equation}
We give a proof in the Supplementary Material.

In the following lemma, we evaluate the variance of GIC at $\theta^*$, similar to Lemma 5 of Song {\it et al.} (2019).


{\bf Lemma 3}:
The $h\times h$ cov-variance matrix for the $h$-dimensional gradient vector of $
GIC(M(\theta))$ is given by
    \begin{equation}
    Var_{p_x}[\nabla_{\theta}GIC(M(\theta^*)]=\Sigma_n(\theta^*)=(\sigma_{i,j}(\theta^*))_{h\times h}
    \end{equation}
    and, for $i,j=1,\cdots,h$,
    \begin{equation}
    \sigma_{i,j}(\theta^*)=\frac{1}{n}\delta_{i,j},
    \end{equation}
 
 where 
    \begin{equation}
    \delta_{i,j}=E_{p_x}[\partial_i W(x,p_{M(\theta^*)})\partial_j W(x,p_{M(\theta^*)})].
    \end{equation}

We give a proof in the Supplementary Material.

{\bf Corollary 6}:
    As $n\rightarrow \infty$, the limit of the left hand side of equation (3.37) is as given in equation (3.38)
    \begin{align}     
    Var_{p_x}[\nabla_{\theta}GIC(M(\theta^*)] &= \frac{1}{n} E_{p_x}[\nabla_{\theta}W(x,p_{M(\theta^*)})\nabla_{\theta}^TW(x,p_{M(\theta^*)})] \\
    &\approx0.
    \end{align}


We give a proof  in the Supplementary Material.

{\bf Proposition 6}:
Under assumptions 1-8 and let $\hat{\theta}_n$ be the maximum GIC estimator defined previously,
we have
\begin{equation}
    \sqrt{n}(\hat{\theta}_n-\theta^*)\xrightarrow{dist} N(0,D^{-1}(\theta^*) \Lambda (\theta^*)D^{-1}(\theta^*)),
\end{equation}
where 
\begin{equation}
   D(\theta^*) =-E_{p_x}[\nabla_{\theta}^2W(x,p_{M(\theta^*)})],
\end{equation}
\begin{equation}
   \Lambda (\theta^*)=E_{p_x}[\nabla_{\theta}W(x,p_{M(\theta^*)})\nabla_{\theta}^TW(x,p_{M(\theta^*)})]
\end{equation}
We give a proof in the Supplementary Material.

\subsubsection{A new model selection method by using the G-cross-entropy}

The model selection based on the Bayes factor is popular. Given data $x=(x_1,\cdots ,x_n)$, the {\it  Bayes factor (BF)} is the ratio of the likelihood  $p(x|M_1)$ of  model $M_1$ to the likelihood  $p(x|M_2)$ of model $M_2$. 
\begin{equation}
BF=\frac{p(x|M_1)}{p(x|M_2)}.
\end{equation}It is considered a measure of the strength of evidence in favor of one model  between two competing ones (Kass and Raftery (1995), Lavine and Schervish (1999)).
By the Bayes formula, we have
\begin{equation}
\frac{p(M_1|x)}{p(M_2|x)}=BF\times \frac{p(M_1)}{p(M_2)}.
\end{equation}
That is
\begin{equation}
\text{ Posterior odds = Bayes factor }\times \text{Prior odds}.
\end{equation}
Although the concept of the Bayes factor is simple and intuitive, depending on the complexity of the model and the hypotheses, the computation of the Bayes factor can be challenging.  
One solution  is to approximate the Bayes factor  by a Laplace approximation (Ibrahim {\it et al.} (2001) 
of the Bayes factor, arriving at the BIC  criteria (Schwarz (1978), Akaike (1978))

\[
BIC=k\times ln(n) - 2ln(\hat{L}),
\]
where $k$ is the dimension of the parameter space, $n$ is the sample size and $\hat{L}$ the maximized value of the likelihood function of model $M$. 
However, closed-form expressions of the 
likelihood are generally unavailable and numerical approximation methods are used such as MCMC (Congdon (2014)). 

In the following, we use the F-cross-entropy to introduce an alternative approach to model selection. 
By Theorem 1,
we have
\begin{equation}
D_F(p||q)=\frac{1}{2}[H_G(p)-H_G(p,q)].
\end{equation}
Therefore, given PDF $p$, minimization with respect to $q$ of the Fisher divergence between $p$ and $q$ is equivalent to maximization  of the G-cross-entropy $H_G(p,q)$.
Thus, when $p$ is the true data PDF and $q$ the model PDF, we can use the G-cross-entropy to develop a tool for model selection.

Let $x_1,\cdots ,x_n$ denote observations from $p_x$, the data PDF. Let M be a model with $p_M$ as its PDF. On using similar arguments as in Hyv{\"a}rinen (2005), it is easy to prove that $GIC(M)$ is a consistent estimate of the G-cross-entropy $H_G(p_x,p_M)$ under assumptions 1-3. Shao {\it et al.} (2019) 
defined a slightly different sample version of $H_G(p_x,p_M)$ and proved that, under some conditions, it is a consistent estimate of $H_G(p_x,p_M)$. 

Now, the G-cross-entropy $H_G(p_x,p_M)$ inspires a criterion for model selection. 


\begin{definition} The following sample sum is called the $GIC_n$ of model $M$:
  \begin{equation}
     GIC_n(M)=\sum_{i=1}^n 
	W(x_i,p_M).
      \end{equation}  
\end{definition}

Similar to Akaike (1985), there is a bias term that needs to be corrected if we wish to use $GIC_n(M)$ for model selection. Suppose we have a collection of candidate {\it parametric} models, say $M_1,\cdots,M_k$ denoted as $M_i(\theta_i), i=1,\cdots,k. $ Note that the different models are associated with different parameters.   
(For convenience, we focus on just one of the $k$ models and drop the suffix $i$ temporarily.)  
Let $x$ denote the observed data. Adopting Akaike's predictive view (1985), let $z$ denote the future data that is independent of $x$ but follows the same probability law as $x$. 

We evaluate the bias $B$ for model $M_n(\hat\theta)$ as follows:
$$B = E_{p_x}[\sum_{j=1}^n 
	W(x_j,p_{M(\hat{\theta})})-nE_{p_z}[W(z,p_{M(\hat{\theta})})]],$$
where $\hat{\theta}=arg \max GIC(M(\theta)).$

 Let $\theta^*$ denote the true $\theta$. 
 We assume that the true data PDF is given by  $p_{M(\theta^*)}$.
Now, writing 
$B=b_1+ b_2+ b_3$, we evaluate, in turn, the $b_i, i=1,2,3$, where 

$b_1 =$
$E_{p_x}[\sum_{j=1}^n 
	W(x_j,p_{M(\hat{\theta})})-\sum_{j=1}^n 
W(x_j,p_{M({\theta}^*)})];$


$b_2=$
 $E_{p_x}[\sum_{j=1}^n 
W(x_j,p_{M({\theta^*})})- nE_{p_z}[W(z,p_{M({\theta^*})})]];$


$b_3=$
 $E_{p_x}[nE_{p_z}[W(z,p_{M({\theta^*})})]-nE_{p_z}
[W(z,p_{M(\hat{\theta})})]].$

Note that 
$b_2$ 
is clearly 0. For 
$b_3$, we expand $W(z,p_{M(\hat{\theta})})$ about $W(z,p_{M({\theta^*})}).$ 
For  
$b_1$, 
we expand $W(x_j,p_{M({\theta}^*)})$ about $W(x_j,p_{M(\hat{\theta})}).$ 
Then, on using similar arguments as in Section 3.4.3 of Konishi and Kitagawa (2008, pp57-59) and Propositions 5 and 6 of this paper, the following proposition is proved.

{\bf Proposition 7}: 
   Under assumptions (1)-(8), a bias-corrected  $FIC_n$, denoted as $FIC_c$,
    for model selection is given by    
    \begin{equation}
        GIC_c(M_i(\hat{\theta_i}))=GIC_n(M_i(\hat{\theta_i}))-B_i,
    \end{equation}
    where $\hat{\theta_i}$ is the maximum GIC estimate of $\theta_i$ and 
    \begin{align}
         B_i&=tr\{\Lambda (\theta^*)D^{-1}(\theta^*)\}\\
    &=-tr\{E_{p_x}[\nabla_{\theta}W(x,p_{M_i(\theta^*)})\nabla_{\theta}^TW(x,p_{M_i(\theta^*)})]E_{p_x}^{-1}[\nabla_{\theta}^2W(x,p_{M_i(\theta^*)})]\}. 
    \end{align}

Note that $W(x_j,p_{M(\theta)})$ mirrors the log likelihood in Konishi and Kitagawa (2008, pp57-59). 

Clearly, $B_i$ will involve the unknown $\theta^*$. 
In practice, we replace the unknown $\theta^*$ by an appropriate estimate.
We denote $GIC_c(M_i(\hat{\theta}_i)) + \hat{B}_i$ by
$\hat{GIC_c}(M_i(\hat{\theta}_i)).$

\begin{example}
Similar to Akaike (1970), consider a stationary autoregressive (AR(p)) model of order $p$ given by
\begin{equation}
x_t=\sum_{i=1}^p a_i x_{t-i} + \epsilon_t,   
\end{equation} 
where the $\epsilon_t$s are independently and identically distributed, with PDF $f_{\epsilon_t} \sim\ N(0,\sigma^2)$. 

Suppose we have a collection of candidate AR(p) models for $p = 0,1,\cdots,L$, $L$ being the maximum possible order. Let $\{x_t, t=1,2, \cdots, N\}$ denote $N$ observations from the AR(p) model. Let $n = N-L.$ Let $\theta=(a_1, a_2, \cdots a_p, \sigma^2)^T$ be the parameter vector of AR(p) model. Assume $f_{x_t}=f_{x_t}{(\theta^*)}=f_{x_t}({x_t|x_{t-1},x_{t-2},\cdots x_{t-p}, \theta^*)},$ where $\theta^*=(a_{1}^{*} , a_{2}^{*}, \cdots a_{p}^{*}, (\sigma^*)^2)^T$ is the true parameter vector in $\Theta$.
Now, 

$$\log f_x = \log f(x_1,x_2,\cdots, x_N) = \sum_{j=1}^N log f(x_j|x_{j-1},x_{j-2},\cdots x_{j-p}) = \sum_{j=1}^N log f_{\epsilon_j}.$$
Using  Definition 2.2 and after some routine calculations, we have
\begin{equation}
    W(x_t,f_{x_t}(\theta))= -\frac{[x_t-(a_1x_{t-1}+\cdots+a_px_{t-p})]^2}{\sigma^4} + \frac{2}{\sigma^2}, t=1,2,\cdots,N,
\end{equation} 

\begin{equation}
    GIC_N(\theta)=\sum_{t=1}^L W(x_t,f_{x_t}(\theta)) + \sum_{t=L+1}^N W(x_t,f_{x_t}(\theta)).
\end{equation} 

Discarding the first sum because $[x_t-(a_1x_{t-1}+\cdots+a_px_{t-p})]^2$ for $t=1, \cdots L,$  are unavailable in the AR(L) model, we have 

\begin{equation}
    GIC_n(\theta)=\sum_{t=L+1}^N W(x_t,f_{x_t}(\theta))=-\frac{\sum_{t=L+1}^N [x_t-(a_1x_{t-1}+\cdots+a_px_{t-p})]^2}{\sigma^4} +\frac{2n}{\sigma^2}.
\end{equation} 

Let $\hat{a}_i$ and $\hat\sigma^2$ be their maximum GIC estimates, which, for the Gaussian case, are identical to the MLEs. Therefore,  
$$\hat\sigma^2=\sum_{t=L+1}^N {\{x_t - \hat{x}_t\}^2 }/n = RSS/n.$$ where $\hat{x}_t = \sum_{i=1}^p \hat{a}_i x_{t-i}$.  

Now,
$$GIC_n(AR(p))=GIC_n(\hat\theta)=\frac{n}{\hat\sigma^2}.$$ 

Since 
\begin{align}
   E_{p_x}[\nabla_{\theta}^2W(x_t,p_{M(\theta^*)})]=
\begin{pmatrix}
  -\frac{2x_{t-1}^2}{(\sigma^*)^4}&  0&\cdots  &0 &0\\
  \vdots &  \vdots&  \ddots &\vdots &\vdots\\
  0&  0&  \cdots &-\frac{2x_{t-p}^2}{(\sigma^*)^4}&0\\
   0&  0&  \cdots &0 &-\frac{2}{(\sigma^*)^6}\\
\end{pmatrix}\nonumber
\end{align}
and
\begin{align}
   E_{p_x}[\nabla_{\theta}W(x_t,p_{M(\theta^*)})\nabla_{\theta}^TW(x_t,p_{M(\theta^*)})]=
\begin{pmatrix}
  \frac{4x_{t-1}^2}{(\sigma^*)^6}&  0&\cdots  &0 &0\\
  \vdots &  \vdots&  \ddots &\vdots &\vdots\\
  0&  0&  \cdots &\frac{4x_{t-p}^2}{(\sigma^*)^6}&0\\
   0&  0&  \cdots &0 &\frac{8}{(\sigma^*)^8}\\
\end{pmatrix}\nonumber,
\end{align}
we have
$$B_p=-tr\{E_{p_x}[\nabla_{\theta}W(x,p_{M(\theta^*)})\nabla_{\theta}^TW(x,p_{M(\theta^*)})]E_{p_x}^{-1}[\nabla_{\theta}^2W(x,p_{M(\theta^*)})]\}=\frac{2(p+2)}{(\sigma^*)^2}$$ 
upon using equation (3.48).

Hence, 
$$ GIC_c(AR(p)) = \frac{n}{\hat\sigma^2} - \frac{2(p+2)}{(\sigma^*)^2}.$$

$${\widehat{GIC_c}}(AR(p)) = \frac{n}{\hat\sigma^2} - \frac{2(p+2)}{\hat\sigma^2}=\frac{n}{RSS/n} \{1 - \frac{2(p+2)}{n}\}.$$

Maximising 
$\widehat{GIC_c}(AR(p))$ 
 is equivalent to minimising $$log (\frac{RSS}{n}) + \frac{2p}{n} + \frac{4}{n},$$
which may be compared with $$log (\frac{RSS}{n}) + \frac{2p}{n} + (1 +log (2\pi))$$ for the AIC.

\end{example}

\subsubsection{Wasserstein distance bounded by Fisher divergence}

The Wasserstein distance is popular in studying  convergence 
of MCMC 
algorithms.
It 
measures the dissimilarity between the target distribution and the distribution obtained from the MCMC samples. 

However, the  Wasserstein distance can be difficult to calculate, especially for large-scale problems. The difficulty arises from the curse of dimensionality and the need to optimize over the space of probability densities. It is further complicated by the fact that it involves finding the minimum cost necessary to move from one distribution to another, which can be computationally intensive, i.e. NP-hard, especially for continuous distributions or high-dimensional spaces (Altschuler and Boix-Adsera (2021)). 

Relationship between the Fisher divergence and Wasserstein distance was
discussed by Huggins {\it et al.} (2018). 
They showed that posterior approximations based on minimizing the Wasserstein distance $D_W$,
rather than the JKL divergence, provide a better error control on moments. They also suggested improved moment estimates by showing that with $\alpha$ a constant  
\begin{equation}
D_W \leq \frac{1}{\alpha}\{ D_F(p||q)\}^{\frac{1}{2}}.
\end{equation}

\section{Gradient-based approach to missing data and discrete data}

\subsection{Various conventional EM algorithms}

\subsubsection{EM algorithm: introduction}

Let $x$ denote the observable data and $z$ the latent variable or the missing data with the following  joint PDF parametrized by  $\theta$: 
\begin{equation}
	p(x,z;\theta)~=~p(x;\theta)p(z|x;\theta).
\end{equation} 
Clearly the complete loglikelihood function decomposes as 
\begin{equation}
	l(\theta|x,z)~=~l(\theta|x)~+~log[p(z|x;\theta)],
\end{equation}
where 
$l(\theta|x)=log[p(x;\theta)]$, the observed data loglikelihood, and $l(\theta|x,z)=log[p(x,z;\theta)].$ Here, $p(z|x;\theta)$, called the
{\it predictive PDF},
plays a central role in the EM algorithm.
Now, $log[p(z|x;\theta)]$ cannot be calculated directly 
because $z$ is unknown.
Instead, we 
average 
equation 
with the predictive PDF $p(z|x;\theta^{(t)})$, where
$\theta^{(t)}$ is a preliminary estimate of the unknown parameter, yielding
\begin{equation}
	Q(\theta |\theta^{(t)})~=~l(\theta |x)~+~H(\theta |\theta^{(t)}), \\where
\end{equation}
\begin{equation}
	Q(\theta |\theta^{(t)})~=~\int_z l(\theta |x,z)p(z|x;\theta^{(t)}) dz~=~E[l(\theta |x, z)|x;\theta^{(t)}],
\end{equation}

\begin{equation}
	H(\theta | \theta^{(t)})=\int_z ~log[p(z|x;\theta)] p(z|x;\theta^{(t)})dz=E[[log[p(z|x;\theta)]|x;\theta^{(t)}],
\end{equation} 
and $ \int l(\theta |x)p(z|x,\theta^{(t)})dz = l(\theta|x)\int p(z|x,\theta^{(t)})dz = l(\theta|x).$

As a result, we have the following EM algorithm.
\begin{itemize}
	\item {\it Expectation or E-step}: Given observed data $x$ and estimated parameter $\theta^{(t)}$, calculate 
 $Q(\theta |\theta^{(t)})$ 
 by averaging the complete-data loglikelihood $l(\theta |x,z)$ with 
 $p(z|x;\theta^{(t)})$.
         \item {\it Maximization or M-step}: Maximize $Q(\theta |\theta^{(t)})$ with respect to $\theta$ to obtain $\theta^{(t+1)}$.
\end{itemize}

Thus, the EM algorithm has converted a function of the unknown $z$ into one of unknown $\theta$ through the ingenious conditional expectation in the E-step, after which the M-step follows.

\subsubsection{When expectation function $H(\theta | \theta^{(t)})$ has no analytic formula for the unknown parameter $\theta$}

Often with missing data $z$, we cannot directly evaluate 
$H(\theta |\theta^{(t)}),$ 
because it involves an integral over $z$. This is especially serious when  $x$ and $z$  are of high dimension. However, the following ways are available to address the issue. 

\begin{itemize}
\item {\bf Monte Carlo EM:} Approximate the expectation in $H(\theta |\theta^{(t)})$ by drawing m random samples $z_1,\cdots ,z_m$ of z from 
$p(z|x,\theta^{(t)})$. Then $H(\theta |\theta^{(t)})$ is approximated by the average 
of these $m$ sampled values of $z$. Specifically,
\begin{equation}
	\hat{H}(\theta |\theta^{(t)})=\frac{1}{m}\sum_{i=1}^mlog[p(z_i|x;\theta^{(t)})].
\end{equation}
As $m$ gets large, this Monte Carlo estimate of H converges to the true value. 

\item {\bf Stochastic EM}: 
Instead of drawing $m$ random samples as above, we can simply take one random sample, $z_1$ from $p(z|x,\theta^{(t)})$ at each E-step. 
Specifically,
\begin{equation}
	\hat{H}(\theta |\theta^{(t)})=log[p(z_1|x;\theta^{(t)})].
\end{equation}
The stochastic nature of this algorithm leads to additional variability, but it still converges in probability to the MLE. 
\item {\bf Bayesian Variational EM}: We postulate a simpler family of distributions $q(z)$ to approximate the true posterior $p(z|x,\theta^{(t)})$. The E-step then minimizes the Kullback-Leibler divergence between q and p. This leads to a tractable lower bound 
on the Shannon entropy $H$. Specifically, 
the E-step starts with  minimizing over $q$
\begin{equation}
	KL(q || p)=\int_z q(z)log[\frac{q(z)}{p(z|x,\theta^{(t)})}]dz.
\end{equation}

\item {\bf MCEM}: Combine the Monte Carlo E-step above with the EM algorithm. This avoids nested simulation loops and is computationally more stable.
\end{itemize}

\subsection{A gradient-based  approach to latent variables and missing data}

Let $p_{data}(x,z)$ be the
full data PDF of $(x,z)$ where $x$ is the observable data in $R^d$ and $z$ the missing data in $R^m$.

Denote the gradient of $log p(x,z,\theta)$ with respect to x and z by  
\begin{align}
	s (x ,z,\theta)&= \begin{bmatrix}
		\nabla_{x} \log ~p(x ,z,\theta) \\
		\nabla_{z} \log ~p(x ,z,\theta) 
	\end{bmatrix}
	= \nabla_{x,z} \log ~p(x ,z,\theta).
\end{align}

Since
\begin{equation}
	\log p(x,z,\theta)=\log p(x,\theta)~+~\log p(z|x,\theta),
\end{equation}
where $p(z|x,\theta)$ is conditional PDF of $z$ given $x$, we have
\begin{align}
	s (x ,z,\theta)&= \begin{bmatrix}
		\nabla_{x} \log p(x, \theta)~+~\nabla_{x} \log p(z|x, \theta)  \\
		\nabla_{z} \log p(z|x,\theta) 
	\end{bmatrix}
\end{align}
\begin{align}
	= \begin{bmatrix}
		\nabla_x \log p(x, \theta) \\
		0 
	\end{bmatrix}
	~+~\begin{bmatrix}
		\nabla_x \log p(z|x, \theta)  \\
		\nabla_z\log p(z|x,\theta)  
	\end{bmatrix}~=~\begin{bmatrix}
		s(x,\theta) \\
		0 
	\end{bmatrix}
	~+~\begin{bmatrix}
		s_x(z|x,\theta) \\
		s_z(z|x,\theta)
	\end{bmatrix}.
\end{align}
Now, the Fisher divergence with the missing $z$ is given by 
\[
J(\theta)~=~\frac{1}{2}E_{p_x}||s(x,\theta)~-~s(x)||^2~+~\frac{1}{2}E_{p_{(x,z)}}||s_x(z|x,\theta)~-~s_x(z|x)||^2
\]
\begin{equation}
	~+~\frac{1}{2}E_{p(x, z)}||s_z(z|x,\theta)~-~s_z(z|x)||^2.
\end{equation}
For $J(\theta)$, we 
define three expectation terms by mimicking the E-step of the EM algorithm:
\begin{equation}
	G_x(\theta | \theta^{(t)}) = \frac{1}{2}E_{p_x}||s(x,\theta)~-~s(x,\theta^{(t)})~||^2,
\end{equation}
\begin{equation}
	H_x(\theta | \theta^{(t)}) = \frac{1}{2}E_{p(x, z)}||s_x(z|x,\theta)~-~s_x(z|x,\theta^{(t)})||^2, 
\end{equation}
and
\begin{equation}
	H_z(\theta | \theta^{(t)}) = \frac{1}{2}E_{p(x, z)}||s_z(z|x,\theta)~-~s_z(z|x,\theta^{(t)})||^2. 
\end{equation}

We can re-formulae the  Fisher divergence equation 
as  
\begin{equation}
	Q_{x,z}(\theta | \theta^{(t)})~=~G_x(\theta | \theta^{(t)})~+~H_x(\theta | \theta^{(t)})~+~H_z(\theta | \theta^{(t)}).
\end{equation}

Now, we discuss how to calculate or approximate the above three terms.

(1) $G_x(\theta | \theta^{(t)})$:
Routine calculations with even the simplest one-dimensional $x\sim N(\mu,\sigma^2)$  indicate that 
 \[
	G_x(\theta | \theta^{(t)})=g(\mu,\sigma^2),
	\]
where $g$ is a polynomial function and is cumbersome to handle. Instead, 
 we  approximate $G_x(\theta | \theta^{(t)})$ as follows.  Given 
 independently and identically distributed observations 
 $x=\{x_1,\cdots ,x_n\}$, we use the approximation
 
\begin{equation}
		\bar{G}_x(\theta | \theta^{(t)})=\frac{1}{2n}\sum_{i=1}^n||s(x_i,\theta)~-~s(x_i,\theta^{(t)})~||^2.
	\end{equation}

(2) $H_x(\theta | \theta^{(t)}) = \frac{1}{2}E_{p(x, z)}||s_x(z|x,\theta)~-~s_x(z|x,\theta^{(t)})||^2:$ 
	Using $p(x,z)=p(z|x)p(x),$ we have
	\[
	H_x(\theta | \theta^{(t)}) = \frac{1}{2}\int_x 	[\int_z ||s_x(z|x,\theta)~-~s_x(z|x,\theta^{(t)})||^2p(z|x)dz]p(x)dx
	\]
	\begin{equation}
		=\int_x\frac{1}{2}\int_x D_x(x,\theta,\theta^{(t)})p(x)dx,
	\end{equation}
	where
	\begin{equation}
		D_x(x,\theta,\theta^{(t)})=\int_z ||s_x(z|x,\theta)~-~s_x(z|x,\theta^{(t)})||^2p(z|x)dz.
	\end{equation}
	Then 
 as in the Monte Carlo EM algorithm, we generate 
 a random sample of size m, $z_1,\cdots , z_m$, from 
 $p(z|x;\theta^{(t)})$ at the {\it t-th-step } of the iteration. Hence, we approximate $D_x(x,\theta,\theta^{(t)})$ by
	\begin{equation}
		D_x^{(t)}(x,\theta,\theta^{(t)})=\frac{1}{m}\sum_{j=1}^m ||s_x(z_j|x,\theta)~-~s_x(z_j|x,\theta^{(t)})||^2.
	\end{equation}
Using observed data $x=\{x_1,\cdots ,x_n\}$ and the above approximator, we approximate $H_x(\theta | \theta^{(t)})$ by
	\begin{equation}
		\bar{H}_x(\theta | \theta^{(t)})=\frac{1}{2n}\sum_{i=1}^nD_x^{(t)}(x_i,\theta,\theta^{(t)}).
	\end{equation}

 (3) $H_z(\theta | \theta^{(t)}) = \frac{1}{2}E_{p(x, z)}||s_z(z|x,\theta)~-~s_z(z|,\theta^{(t)})||^2 $:
 Using the same procedure as above, we approximate $H_z(\theta | \theta^{(t)})$ by
	\begin{equation}
		\bar{H}_z(\theta | \theta^{(t)})=\frac{1}{2n}\sum_{i=1}^nD_z^{(t)}(x_i,\theta,\theta^{(t)}).
	\end{equation}
Putting everything together, we have following {\it gradient-based EM-type  algorithm} for missing data or latent variables:
\begin{itemize}
	\item {\bf Step 1:} initialize estimator $\theta^{(0)};$
	\item {\bf E-step:} calculate the expectation term $Q(\theta | \theta^{(t)})$ as follows.
	\begin{equation}
		\bar{Q}(\theta | \theta^{(t)})~=~ \bar{G}_x(\theta | \theta^{(t)})~+~\bar{H}_x(\theta | \theta^{(t)})~+~\bar{H}_z(\theta | \theta^{(t)});
	\end{equation}
	\item {\bf M-step}: estimate $\theta^{(t+1)}$ by minimization, namely
	\begin{equation}
		\theta^{(t+1)}~=~arg \min_{\theta}\{\bar{Q}(\theta | \theta^{(t)})\}.
	\end{equation}
	Here, in order to obtain an optimal $\theta^{(t+1)}$, we can 
 use gradient descent algorithms such as the SGD algorithm.
Note that a gradient descent algorithm incurs
 second derivatives (the Hessian matrix) of the PDF $p(x,z)$, when 
 calculating the gradient vectors $\nabla_x s(x,\theta)$, $\nabla_x s_x(z|x)$, $\nabla_z s_x(z|x)$, $\nabla_x s_z(z|x)$, and $\nabla_z s_z(z|x)$.
\end{itemize}

\begin{example}
To illustrate the calculation of the various score functions that are essential ingredients of the gradient approach, we consider the  multi-variate t-distribution. Here,

\par
\begin{equation}
\begin{aligned}
    \log{[p(x,\theta)]} =& \log{[\Gamma[(\nu+p)/2]]} - \left( \log{[\Gamma[(\nu)/2]]} + (p/2) \log{[\nu \pi]} + (1/2) \log{[|\Sigma|]} \right) \\
    & - \left( \left( \nu + p \right)/2 \right) \log{\left[ 1 + \frac{1}{\nu} \left(x-\mu_x\right)^T \Sigma^{-1} \left(x-\mu_x\right) \right].}
\end{aligned}
\end{equation}
Then the score function for $x$ is 
\begin{equation}
    s(x,\theta) = \nabla_x{\log{[p(x,\theta)]}} = - \frac{(\nu+p)}{\nu + \left(x-\mu_x\right)^T\Sigma^{-1} \left(x-\mu_x\right)} \cdot \Sigma^{-1}\left(x-\mu_x\right).
\end{equation}
\par
Next, the conditional PDF of z, the conditional mean and the conditional variance are given by the following equations
\begin{equation}
\begin{aligned}
    &p(z|x) = \frac{\Gamma[(\nu+p) / 2]}{\Gamma((\nu + p_x) / 2) (\nu + p_x)^{p_z/2} \pi^{p_z/2} |\Tilde{\Sigma}_{z|x}|^{1/2} } \cdot \\
    &\left[ 1 + \frac{1}{(\nu + p_x)} \left( \left(z-\mu_{z|x} \right)^T \tilde{\Sigma}_{z|x}^{-1} \left(z -\mu_{z|x} \right) \right) \right]^{-(\nu+p) / 2},
\end{aligned}
\end{equation}
\begin{equation}
    \mu_{z|x} = E(z|x) = \mu_z + \Sigma_{zx} \Sigma_{x}^{-1} \left( x - \mu_x \right),
\end{equation}
\begin{equation}
    \Sigma_{z|x} = \Sigma_z - \Sigma_{zx} \Sigma_{x}^{-1} \Sigma_{xz}.
\end{equation}
and
\begin{equation}
    \tilde{\Sigma}_{z|x} = \frac{\nu + d_2}{\nu + p_x} \Sigma_{z|x},
\end{equation}
where $d_2 = (x-\mu_x)^T\Sigma_{x}^{-1} (x-\mu_x)$ is the squared Mahalanobis distance of $x$ from $\mu_x$ with scale matrix $\Sigma_x$. 
\par
Taking the  partial derivative of the log-likelihood with respect to $x$, we have the score function $s_x(z|x, \theta)$ 
given by
\begin{equation}
\begin{aligned}
    & s_x(z|x, \theta) = \nabla_x{\log{[p(z|x, \theta)]}} \\
    & = \frac{\nu + p}{ \nu + p_x + \left( z - \mu_{z|x} \right)^T \tilde{\Sigma}_{z|x}^{-1} \left( z - \mu_{z|x} \right)} \left(\Sigma_{zx}\Sigma_x^{-1}\right)^T \tilde{\Sigma}_{z|x}^{-1} \left( z - \mu_{z|x} \right) + \\
    & \frac{\left(\nu+p\right)\left( z - \mu_{z|x} \right)^T \Sigma_{z|x}^{-1} \left( z - \mu_{z|x} \right)}{\left( \nu + d_2 \right)^{2} \left( 1 + \left( \nu + p_x \right) ^{-1} \left( z - \mu_{z|x} \right)^t \tilde{\Sigma}_{z|x}^{-1} \left( z - \mu_{z|x} \right) \right)} \Sigma_x^{-1} \left( x - \mu_x \right) - \frac{\nabla_x{|\tilde{\Sigma}_{z|x}|}}{2|\tilde{\Sigma}_{z|x}|},
\end{aligned}
\end{equation}
where 
\begin{equation}
    \nabla_x{|\tilde{\Sigma}_{z|x}|} = \frac{2p}{\nu + d_2}\nabla_x{d_2} = \frac{2p}{\nu + d_2} \Sigma_x^{-1}\left( x - \mu_x \right).
\end{equation}
\par
Similarly, partial derivative of the log-likelihood with respect to z yields 
\begin{equation}
\begin{aligned}
    & s_z(z|x, \theta) = \nabla_z{\log{[p(z|x, \theta)]}} \\
    & = \frac{ (\nu + p)\tilde{\Sigma}_{z|x}^{-1}\left( z - \mu_{z|x} \right) }{\nu + p_x + \left( z - \mu_{z|x} \right)^T \tilde{\Sigma}_{z|x}^{-1}\left( z - \mu_{z|x} \right)}.
\end{aligned}
\end{equation}
Finally, we consider the special case when 
$\nu$ tends to infinity. Now,
$s_x(z|x,\theta)$ becomes
\begin{equation}
    s_x^{\nu}(z|x,\theta) = \lim_{\nu \to \infty} s_x(z|x,\theta) = \left(\Sigma_{zx}\Sigma_x^{-1}\right)^T \tilde{\Sigma}_{z|x}^{-1} \left( z - \mu_{z|x} \right),
\end{equation}
and $s_z(z|x, \theta)$ becomes
\begin{equation}
    s_z^{\nu}(z|x,\theta) = \lim_{\nu \to \infty} s_z(z|x,\theta) = \tilde{\Sigma}_{z|x}^{-1}\left( z - \mu_{z|x} \right).
\end{equation}
Clearly, they 
become the score functions of a multi-variate Gaussian PDF. In particular, setting $p_x=1,p_z=1$, we have 
\begin{center}
    $\Sigma =  \begin{pmatrix}
  \sigma_x^2    & \rho \sigma_x \sigma_z \\
  \rho \sigma_x \sigma_z & \Sigma_z^2
\end{pmatrix} $,
\end{center} 
\begin{equation}
    s_x^{\nu}(z|x,\theta) = \rho \frac{\sigma_z}{\sigma_x} \cdot \frac{\left( z - \mu_{z|x} \right)}{\sigma_{z|x}^2},
\end{equation}
and \begin{equation}
        s_z^{\nu}(z|x,\theta) = \frac{\left( z - \mu_{z|x} \right)}{\sigma_{z|x}^2},
\end{equation}
which are the corresponding equations for a bivariate Gaussian PDF.

\end{example}

\subsection{A gradient-based approach to discrete data}

The fact that the probability mass function (PMF) of discrete data does not have gradient poses new challenges. Recently, Grathwohl {\it et al.} (2021) introduced an ingenious device to bridge discrete data with gradient information. Specifically, within an energy-based model framework, they used gradients of the likelihood function with respect to its discrete inputs, and developed  
a general and scalable approximate sampling strategy for 
discrete data.

 Their  setup is 
 an unnormalized PMF 
 \begin{equation}
     p(x) = \frac{1}{Z}exp(f(x)),
 \end{equation} 
 where $Z=\sum_x exp(f(x))$ is the normalizing constant and their attention is restricted to D -dimensional binary data $x\in \{0,1\}^D$ and categorical data $x\in \{0,1,\cdots ,K \}^D$ because all finite-dimensional discrete distributions are embedded
in this way.  
 
 The main contributions of their paper are as follows.
 \begin{itemize}
     \item They have constructed efficient proposal PMFs for discrete discrete data with the new Gibbs-With-Gradients sampling approach that exploits gradient information.
\item Across diverse distributions, their approach beats baseline samplers that do not exploit structure, as well as many  that do exploit hand-coded structure.

\item Their approach can produce 
creatively unconstrained deep energy-based models to be trained on high-dimensional discrete data like images, bettering other deep generative models.

 \end{itemize}
Adopting their energy-based model, we can extend our F-entropy to a discrete random variable $x$. 
 \begin{definition}
     For a discrete random variable $x$, 
the F-entropy of $x$ is defined as
     \begin{equation}
         H_G(x)=E_{p(x)}||\nabla_x f(x)||^2.
     \end{equation}
 \end{definition}
As before, we can define F-cross-entropy, F-conditional entropy and others. We leave further exploration of the above notions to a future project. 

\section{Conclusion}

In this paper, after a bird's-eye view of some 
currently active areas of generative statistics,  we focus on providing a rigorous underpinning to the powerful gradient-based approach that is popular in generative modelling among machine learners by introducing the F-entropy, a new notion of entropy based on the Fisher divergence. The F-entropy, which is of independent interest, parallels the Shannon entropy of conventional statistics that is based on the Jeffreys-Kullback-Leibler divergence but, unlike the latter, it focuses on the information provided by the gradient vector field. The gradient-based approach enjoys various advantages, which include, among others, the freedom from intractable partition function 
We have enlarged the list with a gradient-based algorithm for 
missing data.
 
Finally, we firmly believe that the underpinning  by the F-entropy can and does help the gradient-based approach expand its scope and reap further benefits. We have given an example: 
the G-entropy leads quite naturally to a useful information criterion, the GIC, for generative model selection, thereby highlighting an interesting symmetry with the Shannan entropy approach that created the widely used AIC in the area of statistical model selection.

As for possible future developments, we report how the gradient-based approach has handled Baysian variational inference. A future project is to investigate if and how the G-entropy can help improve its efficacy. As another, 
 we include a brief discussion on the gradient-based approach to discrete distributions and outline possible future developments  based on the G-entropy.

\section{Acknowledgments} 

This work is supported in part by funds from the National Science Foundation (NSF: \# 1636933 and \# 1920920) and Key Laboratory of Random Complex Structures and Data Science, Chinese Academy of Sciences. We thank Miss Rong Bian of the Academy of Mathematics and System Science, Chinese Academy of Sciences, for her assistance.

\section{Bibliography}

\begin{itemize}
\item Akaike, H. (1970). Statistical predictor identification. {\it Ann. Inst. Statist. Math, Part A}, {\bf 22}, 203-217. 

\item Akaike, H. (1978). A Bayesian analysis of the minimum AIC procedure. {\it Ann. Inst. Statist. Math, Part A}, {\bf 30}, 9-14.

\item Akaike, H. (1985).  Prediction and Entropy. {\it A celebration of statistics, The ISI Centenary Volume, edited by Atkinson, A.C. \&  Fienberg, S.E., Springer-Verlag, New York}, 1-24.

\item Altschuler, J.M. \& Boix-Adsera, E. (2021). Wasserstein barycenters are NP-hard to compute. {\it arXiv:2101.01100} 

\item Barron, A.R. (1986). Entropy and the central limit theorem. {\it Annals of Probability}, {\bf 14}, 336–342. 




\item Congdon, P. (2014). Estimating model probabilities or marginal likelihoods in practice. {\it Applied Bayesian Modelling (2nd ed.), Wiley}. 38–40. 

\item Cranmer, K., Brehmer, J. and Louppe, G. (2020). The frontier of simulation-based inference. {\it arXiv:1911.01429}.









\item Grathwohl, W. Swersky, K.,Hashemi, M., Duvenaud, D and Maddison, C.J. (2021). Oops I Took A Gradient: Scalable Sampling for Discrete Distributions
{\it Proc. 38th Inter. Conf. on Machine Learning}, {\bf PMLR 139}.



\item Huggins, J. H., Campbell, T., Kasprzak, M., \& Broderick, T. (2018). Practical bounds on the error of Bayesian posterior approximations: A non-asymptotic approach. {\it arXiv preprint arXiv:1809.09505}.

\item Hyvärinen, A. (2005). Estimation of Non-Normalized Statistical Models by Score Matching. {\it J Machine Learning Research}, {\bf 6}, 695–709.

\item Ibrahim, J.G., Chen, \& M, Sinha, D. (2001) Model Comparison. {\it Bayesian Survival Analysis, Springer} 246–254.

\item Kass, R.E. \& Raftery, A.E. (1995) Bayes Factor.{\it J Amer. Statist. Assoc.}, {\bf 90}, 773-795




\item Kingma, D.P. \& Welling, M. (2022). Auto-Encoding Variational Bayes. {\it arXiv:132.6114}. 

\item Konishi, S. and Kitagawa, G. (2008). {\it Information Criteria and Statistical Modeling}. Springer:New York, USA.


\item  Lavine, M. and Schervish, M.J. (1999). Bayes Factors: What They Are and What They Are Not. {\it The Amer. Statistician}, {\bf  53} ,119-122.

\item Liu, Q., Lee, J. and Jordan, M. (2016). A kernelized stein discrepancy for goodness-of-fit tests. {\it ICML}, 276–284.



\item  Lyu, S.(2009). Interpretation and generalization of score matching. {\it Proceedings of 25th conference on Uncertainty in Artificial Intelligence. (UAI*09)} Montreal QC Canada, 18-21 June 2009. $http://www.cs.mcgill.ca/ uai2009/$ Pages 359-366.




\item Park, S., Serpedin, E., and Qaraqe, K. (2012). On the equivalence between Stein and de Bruijn identities. {\it IEEE Transactions on Information Theory}, {\bf 58} 7045–7067. 







\item Schwarz, G (1978).  Estimating the dimension of a model. {\it Ann. Statist.},  {\bf 6}, 461–464. 

\item Shao, S.,  Jacoba, P.E., Ding, J. \& Tarokh, V. (2019). Bayesian Model Comparison with the Hyvärinen Score: Computation and Consistency. {\it J. Amer. Statist. Assoc. }, {\bf 114}, 1826-1837.


\item Song, Y. and  Ermon, S. (2019). Generative Modeling by Estimating Gradients of the Data Distribution. {\it Advances in Neural Information Processing Systems},  11895--11907. 

\item Song, Y.,  Garg,S. , Shi, J., and Ermon, S. (2019). Sliced Score Matching: A Scalable Approach to Density and Score Estimation. {\it arXiv:1905.07088}.



\item Stein, C. M. (1981). Estimation of the mean of a multivariate normal distribution.  {\it Ann. Statist.},  {\bf 9},  1135–1151.

\item Sueli, I.R., Costa, A., Sandra A. Santos, S.A., and Strapasson, J.E. (2015). Fisher information distance: A geometrical reading. {\it Discrete Appl. Math.}, {\bf 197}, 59-69.



\item Welling, M. and Teh, Y.W. (2011). Bayesian learning via stochastic gradient Langevin dynamics.{\it Proc. 28th international conference on machine learning}, {\bf ICML-11}, 681–688.


$https://en.wikipedia.org/wiki/Prompt\_engineering$


\item Yang, Y., Martin, R. and Bondell, H. (2019). Variational approximations using Fisher divergence. {\it arXiv:1905.05284}.


\end{itemize}

\section*{Supplementary Material}
The supplemental document contains proofs not given in the main text.

\section{proofs of Theorems}

(1) Proof of Theorem 1 
\begin{proof} 		\[
		E_p||\nabla_x \log p(x)~-~\nabla_x \log q(x)||^2~=~E_p||\nabla_x \log p(x)||^2~+~E_p||\nabla_x \log q(x)||^2
		\]
		\[
		-2E_p[[\nabla_x \log p(x)]^T[\nabla_x \log q(x)]]
		\]
		Now,
		\[
		E_p[[\nabla_x \log p(x)]^T[\nabla_x \log q(x)]]=\sum_{i=1}^dE_p[\frac{\partial \log p(x)}{\partial x_i}\times  \frac{\partial \log q(x)}{\partial x_i}]
		\]
		\[
		=\sum_{i=1}^d\int_{R^d}\frac{\partial \log p(x_1,\cdots,x_d)}{\partial x_i} \frac{\partial \log q(x_1,\cdots ,x_d)}{\partial x_i}p(x_1,\cdots ,x_d)dx_1\cdots dx_d.
		\]
		For each $i,~i=1,\cdots ,d$, , 
		the multiple integral over $R^d$ is given by
		
		\[
		\int_{x_1,\cdots,x_{i-1},x_{i+1},\cdots ,x_d}\{\int_{x_i}\frac{\partial \log p(x_1,\cdots,x_d)}{\partial x_i} \frac{\partial \log q(x_1,\cdots ,x_d)}{\partial x_i}
		\]
		\[
		p(x_1,\cdots ,x_d) dx_i\}dx_1,\cdots,dx_{i-1},dx_{i+1},\cdots ,dx_d.
		\]
		
		Now,
		\[
		\int_{x_i}\frac{\partial \log p(x_1,\cdots,x_d)}{\partial x_i} \frac{\partial \log q(x_1,\cdots ,x_d)}{\partial x_i}p(x_1,\cdots ,x_d) dx_i
		\]
		\[ 
		=\int_{x_i}\frac{\partial  p(x_1,\cdots,x_d)}{\partial x_i} \frac{\partial \log q(x_1,\cdots ,x_d)}{\partial x_i} dx_i
		\]
		\[ 
		= \int_{x_i}\frac{\partial \log q(x_1,\cdots ,x_d)}{\partial x_i}
		p(x_1,\cdots,x_{i-1},dx_i,x_{i+1},\cdots ,x_d) 
		\]
		\[
		=\frac{\partial \log q(x_1,\cdots ,x_i,\cdots ,x_d)}{\partial x_i}
		p(x_1,\cdots,x_{i-1},x_i,x_{i+1},\cdots ,x_d)|_{x_i=-\infty}^{x_i=+\infty}
		\]
		\[
		-\int_{x_i}\frac{\partial^2 \log q(x_1,\cdots ,x_i,\cdots ,x_d)}{\partial^2 x_i}
		p(x_1,\cdots,x_{i-1},x_i,x_{i+1},\cdots ,x_d)dx_1\cdots dx_d,
		\]
		by using the integration by parts.
		
		Next, by assumption 3, the first term above is equal to zero. Therefore
		\[
		\int_{x_1,\cdots,x_{i-1},x_{i+1},\cdots ,x_d}\{\int_{x_i}\frac{\partial \log p(x_1,\cdots,x_d)}{\partial x_i} \frac{\partial \log q(x_1,\cdots ,x_d)}{\partial x_i}
		\]
		\[
		p(x_1,\cdots ,x_d) dx_i\}dx_1\cdots,dx_{i-1}dx_{i+1}\cdots ,dx_d
		\]
		\[
		=-\int_{x_1,\cdots,x_{i-1},x_{i+1},\cdots ,x_d}\{\int_{x_i} \frac{\partial^2 \log q(x_1,\cdots ,x_d)}{\partial^2 x_i}
		\]
		\[
		p(x_1,\cdots ,x_d) dx_i\}dx_1\cdots,dx_{i-1}dx_{i+1}\cdots ,dx_d
		\]
		\[
		=-\int_{x_1,\cdots,x_{i-1},x_i,x_{i+1},\cdots ,x_d}\frac{\partial^2 \log q(x_1,\cdots ,x_d)}{\partial^2 x_i}p(x_1,\cdots ,x_d) \}
		\]
		\[
		dx_1\cdots,dx_i \cdots dx_d
		\]
		\[
		=-E_p[\frac{\partial^2 q(x)}{\partial^2 x_i}]
		\]
		Therefore we obtain
		\[
		E_p[[\nabla_x \log p(x)]^T[\nabla_x \log q(x)]]=-\sum_{i=1}^dE_p[\frac{\partial^2 q(x)}{\partial^2 x_i}]
		\]
		Hence
		\[
		E_p||\nabla_x \log p(x)~-~\nabla_x \log q(x)||^2
		\]
		\[
		=E_p||\nabla_x \log p(x)||^2~+~E_p||\nabla_x \log q(x)||^2~+~2\sum_{i=1}^dE_p[\frac{\partial^2 q(x)}{\partial^2 x_i}]
		\] 
		\[		
		=E_p||\nabla_x \log p(x)||^2~-~E_p\{W(x,q)\}.
		\]
		The theorem is proved. 
\end{proof}

(2) Proof of Theorem 2 
\begin{proof}
Let the dimension of x be d and the dimension of y be e. 
By the definition of the G-divergence 
and Theorem 1, we have
\[
D_F(p_{(x,y)}||p_xp_y)=\frac{1}{2}\{-E_{p_{(X,Y)}}[W(\text{\small (x,y)},~p_xp_y)]+H_G(p_{(x,Yy)})\}
\]
\begin{equation}
	=\frac{1}{2}\{E_{p_{(x,y)}}||\nabla_{x,y}[\log \{p_x(x) p_y(y)\}]||^2~+~2E_{p_{(x,y)}}[\Delta_{x,y}\log \{p_x(x)p_y(y)\}]~+~H_G(p_{(x,y)})\}.
\end{equation}
Now, the term
\[
E_{p_{(x,y)}}||\nabla_{x,y}[\log \{p_x(x) p_y(y)\}]||^2=E_{p_{(x,y)}}||\nabla_{x,y}[\log p_x(x) + \log p_y(y)]||^2.
\]
Since $\nabla_{x,y}\log p_x(x)$ is a $d\times e$ matrix given by
\begin{align*}		
	\nabla_{x,y}\log p_x(x) = \begin{bmatrix}
		\nabla_x \log p_x(x) \vspace{0.3cm}\\
		0 
	\end{bmatrix}_{d\times e}
\end{align*}
and $\nabla_{x,y}\log p_y(y)$ is a $d\times e$ matrix
given by
\begin{align*}		
	\nabla_{x,y}\log p_y(y) = \begin{bmatrix}
		0 \vspace{0.3cm}\\
		\nabla_y \log p_y(y) 
	\end{bmatrix}_{d\times e},
\end{align*}
therefore
\[
E_{p_{(x,y)}}||\nabla_x\log p_x(x) + \nabla_y\log p_y(y))]||^2_{d\times e}
\]
\[
=E_{p_{(x,y)}}||\nabla_x\log p_x(x)||^2_d+E_{p_{(x,y)}}||\nabla_y\log p_y(y)||^2_e
\]
\begin{equation}
	=H_G(p_x)+H_G(p_y)=H_G(x)+H_G(y)
\end{equation}
by using Proposition 4. 
Next, for the second term in equation (8.1), 
\[
E_{p_{(x,y)}}[\Delta_{x,y}\log p_x(x)p_y(y)]=E_{p_{(x,y)}}[\Delta_{x,y}[\log p_x(x)+\log p_y(y)]]
\]
\[
=E_{p_{(x,y)}}[\Delta_{x,y}\log p_x(x)]+E_{p_{(x,y)}}[\Delta_{x,y}\log p_y(y)]
\]
\[
=E_{p_{(x,y)}}[\Delta_x\log p_x(x)]+E_{p_{(x,y)}}[\Delta_y\log p_y(y)]
\]
since
\[
E_{p_{(x,y)}}[\Delta_x\log p_x(x)]=\int_x\int_y \{\Delta_x\log p_x(x)\}p_{(x,y)}(x,y)dxdy
\]
\[
=\int_x \Delta_x\log p_x(x)[\int_y p_{(x,y)}(x,y)dy]dx
\]
\[
=\int_x [\Delta_x\log p_x(x)]p_x(x)dx=E_{p_x}[\Delta_x\log p_x(x)]
\]
\[
=-E_{p_x}||\nabla_x \log p_x(x)||^2=-H_G(x).
\]
Similarly we have
\[
E_{p_{(x,y)}}[\Delta_y\log p_y(y)]=-H_G(y).
\]
Putting the two together, we obtain
\[
E_{p_{(x,y)}}[W( (x,y),~p_xp_y)]
\]
\[
=-[H_G(x)+H_G(y)]+2[H_G(x)+H_G(y)]=H_G(x)+H_G(y).
\]
Finally we obtain
\[
D_F(p_{(x,y)}||p_xp_y)=\frac{1}{2}\{-H_G(x)-H_G(y)+H_G(x,y)\}=\frac{1}{2}I_G(x,y).
\]
Hence, the theorem is proved. 
\end{proof}

(3) Proof of Theorem 3
\begin{proof}
Let the dimension of x be d and the dimension of y be e.

For brevity, we denote the joint PDF of  $(x,y)$ as $p(x,y)$ and the marginal PDF of x as $p(x)$. 
By the definition,
we have
\[
H_G(y|x)=E_{p(x,y)}[W((x,y),\frac{p(x,y)}{p(x)})]
\]
	$$=-E_{p(x,y)}||\nabla_{x,y}\log(\frac{p(x,y)}{p(x)})||^2-2E_{p(x,y)}[\Delta_{x,y}\log (\frac{p(x,y)}{p(x)})]. \hspace{1cm}   (106)$$

For the first term,
\[
E_{p(x,y)}||\nabla_{x,y}\log(\frac{p(x,y)}{p(x)})||^2=E_{p(x,y)}||\nabla_{x,y}\log p(x,y)-\nabla_{x,y}\log p(x)||^2
\]
\[
=E_{p(x,y)}|| \begin{bmatrix}
	\frac{\partial \log p(x,y)}{\partial x} \\
	\frac{\partial \log p(x,y)}{\partial y}
\end{bmatrix}_{d\times e} 
-	
\begin{bmatrix}
	\frac{\partial \log p(x)}{\partial x} \\
	0
\end{bmatrix}_{d\times e}
||^2
\]

	$$=E_{p(x,y)}||\nabla_x \log p(x,y)-\nabla_x \log p(x)||^2_d+E_{p(x,y)}||\nabla_y \log p(x,y)||^2_e. \hspace{1cm} (107)$$
Now, \[
E_{p(x,y)}||\nabla_x \log p(x,y)-\nabla_x \log p(x)||^2_d=E_{p(x,y)}||\nabla_x \log p(x,y)||^2_d 
\]
\[
+ E_{p(x,y)}||\nabla_x \log p(x)||^2_d
-2E_{p_{(x,y)}}[[\nabla_x\log p(x,y)]^T[\nabla_x \log p(x)]].
\]
First, 
\[
E_{p(x,y)}||\nabla_x \log p(x,y)||^2_d + E_{p(x,y)}||\nabla_y \log p(x,y)||^2_e = E_{p(x,y)}||\nabla_{x,y} \log p(x,y)||^2_{d\times e}
\]
\[
=H_G(p(x,y))=H_G(x,y).
\]
Next, by Proposition 4, 
we have
\[
E_{p(x,y)}||\nabla_x \log p(x)||^2_d = H_G(x).
\]
Lastly, by Lemma 1 
and Proposition 1,
we have

\[
-2E_{p_{(x,y)}}[[\nabla_x\log p(x,y)]^T[\nabla_x \log p(x)]]=2E_{p_x}[\Delta_x\log p(x)]=-2H_G(x).
\]
Putting the above three result together, we obtain a result for equation (107)
by
	$$E_{p(x,y)}||\nabla_{x,y}\log(\frac{p(x,y)}{p(x)})||^2=H_G(x,y)+H_G(x)-2H_G(x)=H_G(x,y)-H_G(x).\hspace{1cm}(108)$$
For the second term of equation (106), 
we have
\[
E_{p(x,y)}[\Delta_{x,y} \log (\frac{p(x,y)}{p(x)})]=E_{p(x,y)}[\Delta_{x,y} \log p(x,y)]-E_{p(x,y)}[\Delta_{x,y} \log p(x)].
\]
Clearly
\[
E_{p(x,y)}[\Delta_{x,y} \log p(x)]=E_{p(x,y)}[\Delta_x \log p(x)]
\]
\[
=\int_x \Delta_x \log p(x) \{\int_y p(x,y)dy\}dx=\int_x \Delta_x \log p(x) p(x)dx.
\]

\[
=E_{p(x)}[\Delta_x \log p(x)]=-H_G(x),
\]
by  Proposition 1.
Therefore, the second term of equation (106)
is equal to
	$$E_{p(x,y)}[\Delta_{x,y} \log (\frac{p(x,y)}{p(x)})]=-H_G(x,y)+H_G(x).\hspace{1cm} (109)$$
Finally, by substituting (108) and (109)
we obtain
\[
H_G(y|x)=-[H_G(x,y)-H_G(x)]-2[-H_G(x,y)+H_G(x)]=H_G(x,y)-H_G(x).
\]
The theorem is proved. 
\end{proof}

\section{proofs of propositions}

(1) Proof of Proposition 1 
 \begin{proof}
\[
E_p[\nabla_x\log p(X)]=\int_x \begin{bmatrix}
\frac{\partial \log p(x)}{\partial x_1} \\
\vdots\\
\frac{\partial \log p(x)}{\partial x_d}
\end{bmatrix}_{d}
p(x)dx.
\]
For $i=1,\cdots ,d$, 
\[
\int_x \frac{\partial \log p(x)}{\partial x_i} p(x) dx=\int_{x_1,\cdots,x_{i-1},x_{i+1},\cdots ,x_d}\{\int_{x_i}\frac{\partial \log p(x)}{\partial x_i} p(x)dx_i\}
\]
\[
dx_1\cdots dx_{i-1}dx_{i+1}\cdots dx_d
\]
\[
=\int_{x_1,\cdots,x_{i-1},x_{i+1},\cdots ,x_d}\{\int_{x_i}\frac{\partial p(x)}{\partial x_i} dx_i\}dx_1\cdots dx_{i-1}dx_{i+1}\cdots dx_d
\]
\[
=\int_{x_1,\cdots,x_{i-1},x_{i+1},\cdots ,x_d}\{p(x_1,\cdots ,+\infty ,\cdots ,x_d)-p(x_1,\cdots ,-\infty ,\cdots ,x_d)\}
\]
\[
dx_1\cdots dx_{i-1}dx_{i+1}\cdots dx_d,
\]

which is equal to zero by assumption 3. Thus, 
\[
E_p[\nabla_x\log p(x)]~=0.
\]
By simply using the integration by parts, we can prove
\[
E_p[\frac{\partial \log p(x)}{\partial x_i}]^2~=~-E_p[\frac{\partial^2 \log p(x)}{\partial^2 x_i}].
\]
Part (iii) is obvious. Thus,  Proposition 1 is proved. 
\end{proof}

(2) Proof of Proposition 5 
\begin{proof}
    From equation (3) in Song {\it et al.} (2019), it is clear that their objective function is equal to our GIC by
    \[
    \hat{J}(\theta ,x_1^n)=-\frac{1}{2}GIC({M(\theta)}).
    \]
    
    The sample objective function of their sliced score-matching estimation is equation (7) in their paper:
    \[
    \hat{J}(\theta, x_1^n, v_{11}^{nm})=\frac{1}{n}\frac{1}{m}\sum_{i=1}^n\sum_{j=1}^m[v_{i,j}^T\nabla_x^2 \log p_{M(\theta}(x_i) v_{i,j}+\frac{1}{2}(v_{i,j}^T\nabla_x \log p_{M(\theta}(x_i))^2],
    \]
    where $v_{i,j}$ is a d-dimensional projection vector and $d$ is the dimension of the data vector $x$.

    Now, for $m=d$ and for each $i$, we 
    define special projection vectors $v_1,\cdots ,v_m$, of which 
    $v_j=(v_{k,j})$ such that
    \[
    v_{k,j} = \begin{cases}
    1 & if ~k=j\\
    0 & otherwise ~ k \neq j.
    \end{cases}
    \]
    By using these 
    special projection vectors, we obtain
    \[
    \hat{J}(\theta, x_1^n, v_{11}^{nm})=-\frac{1}{2d}\times\frac{1}{n}\sum_{i=1}^nW(x_i,p_{M(\theta)})=-\frac{1}{2d}GIC(M(\theta)).
    \]
    Then, by using Theorems 1 and 2 of Song {\it et al.} (2019), the proof is complete.
    
\end{proof}

(3) Proof of Proposition 6 

\begin{proof}
To simplify notations, we define an average operator, $P_n$, for data sample $x_1,\cdots ,x_n$ applied to any function $g(x,\cdot)$ by
\[
P_n[g(x)]=\frac{1}{n}\sum_{i=1}^ng(x_i,\cdot).
\]
Therefore, we have that
\[
GIC(M(\theta))=P_n[W(x,p_{M(\theta)})].
\]
Given that $\hat{\theta}_n$ is the maximum GIC estimator, we know that
\[
\nabla_{\theta}GIC(M(\hat{\theta}_n)) = 0.
\]
On the other hand, using Taylor expansion, we have
\[
\nabla_{\theta}GIC(M(\hat{\theta}_n))=P_n[\nabla_{\theta}W(x,p_{M(\hat{\theta}_n)})]
\]
\[
=P_n[\nabla_{\theta}W(x,p_{M(\theta^*)})]+P_n[\nabla^2_{\theta}W(x,p_{M(\tilde{\theta}_n)})](\hat{\theta}_n-\theta^*)
\]
$$
=P_n[\nabla_{\theta}W(x,p_{M(\theta^*)})]+P_n[\nabla^2_{\theta}W(x,p_{M(\theta^*)})+\delta(M(\tilde{\theta}_n))](\hat{\theta}_n-\theta^*),\hspace{0.5cm}(110)$$
where $\tilde{\theta}_n$ is a point between $\theta^*$ and $\hat{\theta}_n$. Since $\hat{\theta}_n$ is a consistent estimate of $\theta^*$,then so is $\tilde{\theta}_n$.

From Lemma 2
and by a Taylor expansion of vector-valued functions, we have that
\[
||\delta(M(\tilde{\theta}_n))||_F\leq L(x)||\tilde{\theta}_n - \theta^*||_2.
\]
By the law of large numbers,
\[
P_n[\nabla^2_{\theta}W(x,p_{M(\theta^*)})]=E_{p_x}[\nabla^2_{\theta}W(x,p_{M(\theta^*)})]+o_p(1),
\]
and
\[
||P_n[\delta(M(\tilde{\theta}_n))]||_F\leq E_{p_x}L(x)||\tilde{\theta}_n - \theta^*||_2+o_p(1)=o_p(1)+o_p(1)=o_p(1),
\]
where  $E_{p_x}L(x) \leq \sqrt{E_{p_x}L^2(x)} < \infty$ by Lemma 2 
and the consistency of $\hat{\theta}_n$ (Proposition 5).

Now, returning to equation (110),
we have
\[
0=P_n[\nabla_{\theta}W(x,p_{M(\theta^*)})]+P_n[\nabla^2_{\theta}W(x,p_{M(\theta^*)})+o_p(1)](\hat{\theta}_n-\theta^*)
\]
\[
\iff -P_n[\nabla^2_{\theta}W(x,p_{M(\theta^*)})+o_p(1)]\sqrt{n}(\hat{\theta}_n-\theta^*) =\sqrt{n}P_n[\nabla_{\theta}W(x,p_{M(\theta^*)})].
\]

Therefore, using the central limit theorem for {\it i.i.d.} random variables and Lemma 3,
we have
\[
\sqrt{n}P_n[\nabla_{\theta}W(x,p_{M(\theta^*)})]\xrightarrow{dist}N(0,nVar_{p_x}[\nabla_{\theta}GIC(M(\theta^*))]=N(0,\Lambda (\theta^*)), \hspace{1cm} (111)
\]
where 
   $$\Lambda (\theta^*)=E_{p_x}[\nabla_{\theta}W(x,p_{M(\theta^*)})\nabla_{\theta}^TW(x,p_{M(\theta^*)})].$$

Given that
\[
-\nabla_{\theta}^2GIC(M(\theta^*)=-P_n[\nabla_{\theta}^2W(x,p_{M(\theta^*}))] \xrightarrow{p} D(\theta^*),
\]
where 
$$D(\theta^*) =-E_{p_x}[\nabla_{\theta}^2W(x,p_{M(\theta^*)})],$$
then using the well-known Slutsky’s lemma in probability,
 we have
    $$\sqrt{n}(\hat{\theta}_n-\theta^*)\xrightarrow{dist} N(0,D^{-1}(\theta^*) \Lambda (\theta^*)D^{-1}(\theta^*)). \hspace{1cm} (112)$$

\end{proof}

\section{proofs of lemmas}
*******************************************************************

(1) Proof of Lemma 1 

\begin{proof}
			\[
			E_{p_{(x,y)}}[[\nabla_x\log p(x,y)]^T[\nabla_x \log p(x)]]=\sum_{i=1}^d E_{p_{(x,y)}}[\frac{\partial \log p(x,y)}{\partial x_i}\frac{\partial \log p(x)}{\partial x_i}],
			\]
			where we have denoted $p_{(x,y)}(x,y)$ by $p(x,y)$ and $p_x(x)$ by $p(x)$.
			
			For each $i=1,\cdots ,d$ , 
			
			\[
			E_{p_{(x,y)}}[\frac{\partial \log p(x,y)}{\partial x_i}\frac{\partial \log p(x)}{\partial x_i}]=\int_y\{ \int_x [\frac{\partial \log p(x,y)}{\partial x_i}\frac{\partial \log p(x)}{\partial x_i}] p(x,y)dx\}dy
			\]
			\[
			=\int_y\{ \int_x [\frac{\partial p(x,y)}{\partial x_i}\frac{\partial \log p(x)}{\partial x_i}] dx\}dy=\int_y\{ \int_x [\frac{\partial \log p(x)}{\partial x_i}] p(\partial x_i,y)\}dy
			\]

			\[
			=-\int_x\frac{\partial^2 \log p(x)}{\partial^2 x_i}\{\int_y p(x,y)dy\}dx=-\int_x\frac{\partial^2 \log p(x)}{\partial^2 x_i}p(x)dx,
			\]
			after using an integration by parts and Assumption 3.	Therefore
			\[
			E_{p_{(x,y)}}[[\nabla_x\log p(x,y)]^T[\nabla_x \log p(x)]]=-\sum_{i=1}^d \int_x\frac{\partial^2 \log p(x)}{\partial^2 x_i}p(x)dx
			\]
			\[
			=-E_{p(x,y)}[\Delta_x \log p(x)].
			\]
		\end{proof}

(2) Proof of Lemma 2
\begin{proof}

By using the special projection vectors $v_1,\cdots , v_d$ defined as follow
$v_j=(v_{k,j})$ such that
    \[
    v_{k,j} = \begin{cases}
    1 & if ~k=j\\
    0 & otherwise ~ k \neq j.
    \end{cases}
    \]
we obtain
\[
f(\theta;x,v_1^M) = -\frac{1}{2d} GIC({M(\theta)}) \text{ for } M=d,
\]

where $f(\theta;x,v_1^M)$ is as defined in the Appendix B.4 of the Song {\it et al.} (2019). Then by simply using the conclusion of their Lemma 4, the proof is complete.
\end{proof}

(3) Proof of Lemma 3

\begin{proof} 
Here, we adopt a similar strategy as the proof for Lemma 5 in the Song {\it et al.} (2019) and employ the above special projection vectors $v_1,\cdots , v_d$.

     First, since $\theta^*$ is the true parameter of the data PDF, from Corollary 1 we have
\[
\nabla_{\theta}E_{p_x}[W(x,p_{M(\theta^*)})]=0.
\]
Hence we have
    $$E_{p_x}[\nabla_{\theta}GIC(M(\theta^*)]=\nabla_{\theta}E_{p_x}[GIC(M(\theta^*)]=\nabla_{\theta}E_{p_x}[W(x,p_{M(\theta^*)})]=0.\hspace{1cm}(113)$$
Therefore
\[
Var_{p_x}[\nabla_{\theta}GIC(M(\theta^*)] 
=E_{p_x}[[\nabla_{\theta}GIC(M(\theta^*)][\nabla_{\theta}^{T}FIC(M(\theta^*)]]=\Sigma_n(\theta^*)=(\sigma_{i,j}(\theta^*))_{h\times h},
\]
where
\[
\sigma_{i,j}(\theta^*)=E_{p_x}[\partial_i GIC(M(\theta))\partial_j GIC(M(\theta))]
\]
\[
=\frac{1}{n^2}\sum_{k=1}^n\sum_{l=1}^nE_{p_x}[\partial_i W(x_k,p_{M(\theta^*)})\partial_j W(x_l,p_{M(\theta^*)})].
\]

For the random sample 
$x_1,\cdots ,x_n$, 
we have, for $k=l$,
\[
E_{p_x}[\partial_i W(x_k,p_{M(\theta^*)})\partial_j W(x_k,p_{M(\theta^*)})]
\]
\[
=E_{p_x}[\partial_i W(x,p_{M(\theta^*)})\partial_j W(x,p_{M(\theta^*)})] = \delta_{i,j},
\]
and, for $k\neq l$, by independence between $x_k$ and $x_l$
\[
E_{p_x}[\partial_i W(x_k,p_{M(\theta^*)})\partial_j W(x_l,p_{M(\theta^*)})]
\]
\[
=E_{p_x}[\partial_i W(x_k,p_{M(\theta^*)})]E_{p_x}[\partial_j W(x_l,p_{M(\theta^*)})] = \delta_i\times \delta_j=0,
\]
where $\delta_i = E_{p_x}[\partial_i W(x,p_{M(\theta^*)})]$. Therefore we obtain
\[
\sigma_{i,j}(\theta^*)=\frac{1}{n}\delta_{i,j}+\frac{n-1}{n}\delta_i\times \delta_j=\frac{1}{n}\delta_{i,j}.
\]
\end{proof}

\section{proof of corollaries}

(1) Proof of Corollary 6  

\begin{proof}
   By equation (3.36) in Lemma 3, as $n\rightarrow \infty$, limit of $\sigma_{i,j}(\theta^*)$ in equation  (3.35) of the Main Text is
   \[
    \lim_{n\rightarrow \infty} \sigma_{i,j}(\theta^*) =\lim_{n\rightarrow \infty}\frac{1}{n}\delta_{i,j} =\lim_{n\rightarrow \infty}\frac{1}{n}E_{p_x}[\partial_i W(x,p_{M(\theta^*)})\partial_j W(x,p_{M(\theta^*)})] =0
    \]
    Hence, as $n\rightarrow \infty$,  
        $$Var_{p_x}[\nabla_{\theta}GIC(M(\theta^*)]=(\sigma_{i,j}(\theta^*))_{h\times h}\rightarrow 0.\hspace{2cm}(117)$$
    The proof is complete.
\end{proof}

\end{document}